\documentclass[fleqn,usenatbib]{mnras}

\usepackage{newtxtext,newtxmath}

\usepackage[T1]{fontenc}
\usepackage{ae,aecompl}

\usepackage{graphicx}	
\usepackage{amsmath}	
\usepackage{amssymb}	
\usepackage[usenames,dvipsnames]{color}
\usepackage{float}
\usepackage{pdflscape}
\usepackage{soul}
\usepackage[normalem]{ulem}

\newcommand{\pkg}[1]{\texttt{#1}}

\newcommand{\Msun}{\mathrm{M}_\mathrm{\odot}}
\newcommand{\Msunh}{\Msun\, \mathrm{h}^{-1}}
\newcommand{\Gpc}{\mathrm{Gpc}}
\newcommand{\Gpch}{\mathrm{Gpc}\, \mathrm{h}^{-1}}

\newcommand{\kpc}{\mathrm{kpc}}
\newcommand{\Mpc}{\mathrm{Mpc}}
\newcommand{\Msunyr}{\Msun\, \mathrm{yr}^{-1}}
\newcommand{\Mstar}{M_\mathrm{*,gal}}
\newcommand{\Mgas}{M_\mathrm{gas,gal}}
\newcommand{\Mvir}{M_\mathrm{vir}}
\newcommand{\Vvir}{V_\mathrm{vir}}
\newcommand{\Cvir}{C_\mathrm{vir}}

\newcommand{\gasfrac}{f_\mathrm{gas,gal}}
\newcommand{\Omegamnow}{\Omega_\mathrm{m,0}}
\newcommand{\Omegalambnow}{\Omega_\mathrm{\Lambda,0}}
\newcommand{\Gyr}{\mathrm{Gyr}}
\newcommand{\Myr}{\mathrm{Myr}}
\newcommand{\sigmaeight}{\sigma_\mathrm{8}}

\title[Early BCG Assembly]{Rapid early coeval star formation and assembly of the most-massive galaxies in the universe}

\author[D. Rennehan et al.]{
Douglas Rennehan,$^{1}$\thanks{E-mail: douglas.rennehan@gmail.com (DR)}
Arif Babul,$^{1}$
Christopher C. Hayward,$^{2}$
Connor Bottrell,$^{1}$
\newauthor \ Maan H. Hani,$^{1}$\thanks{Vanier Scholar}
and Scott C. Chapman$^{1,3,4,5}$
\\
$^{1}$Department of Physics \& Astronomy, University of Victoria, BC, V8X 4M6, Canada\\
$^{2}$Center for Computational Astrophysics, Flatiron Institute, 162 Fifth Avenue, New York, NY, 10010, USA\\
$^{3}$Department of Physics and Atmospheric Science, Dalhousie University, Halifax, NS, B3H 4R2, Canada\\
$^{4}$NRC Herzberg Astronomy and Astrophysics, 5071 West Saanich Road, Victoria, BC, V9E 2E7, Canada,\\
$^{5}$Department of Physics and Astronomy, University of British Columbia, Vancouver, BC, V6T 1Z1, Canada
}

\date{Accepted XXX. Received YYY; in original form ZZZ}

\pubyear{2019}

\begin{document}
\label{firstpage}
\pagerange{\pageref{firstpage}--\pageref{lastpage}}
\maketitle

\begin{abstract}
   The current consensus on the formation and evolution of the brightest cluster galaxies is that their stellar mass forms early ($z \gtrsim 4$) in separate galaxies that then eventually assemble the main structure at late times ($z \lesssim 1$).  However, advances in observational techniques have led to the discovery of protoclusters out to $z \sim 7$.  If these protoclusters assemble rapidly in the early universe, they should form the brightest cluster galaxies much earlier than suspected by the late-assembly picture.  Using a combination of observationally constrained hydrodynamical and dark-matter-only simulations, we show that the stellar assembly time of a sub-set of brightest cluster galaxies occurs at high redshifts ($z > 3$) rather than at low redshifts ($z < 1$), as is commonly thought.  We find, using isolated non-cosmological hydrodynamical simulations, that highly overdense protoclusters assemble their stellar mass into brightest cluster galaxies within $\sim 1$ $\Gyr$ of evolution -- producing massive blue elliptical galaxies at high redshifts ($z \gtrsim 1.5$).  We argue that there is a downsizing effect on the cluster scale wherein some of the brightest cluster galaxies in the cores of the most-massive clusters assemble earlier than those in lower-mass clusters. In those clusters with $z = 0$ virial mass $\geqslant 5\times 10^{14}$ $\Msun$, we find that $9.8\%$ have their cores assembly early, and a higher fraction of $16.4\%$ in those clusters above $10^{15}$ $\Msun$.  The James Webb Space Telescope will be able to detect and confirm our prediction in the near future, and we discuss the implications to constraining the value of $\sigmaeight$.
\end{abstract}




\section{Introduction}
\label{sec:introduction}

Galaxy clusters are at the peak of the mass-assembly hierarchy in the $\Lambda\mathrm{CDM}$ paradigm and share a common feature of hosting a distinct population of galaxies aptly referred to as the brightest cluster galaxies (BCGs) \citep{Peebles1968, Sandage1976, Tremaine1977}.  The BCGs are ultra-luminous (with $\sim 10 L_{\mathrm{K},\ast}$ where $L_{\mathrm{K},\ast}$ is the characteristic luminosity of the general galaxy population), morphologically spheroidal, have a large spatial extent, and exhibit core stellar velocity dispersions of order $300-400$ $\mathrm{km \, s^{-1}}$  \citep{Lin2004, Pipino2011, Loubser2018}.  The BCGs are in fact the brightest and the most-massive galaxies in the present-day Universe.  Many of the observed characteristics of the BCGs seem to scale with the properties of the hosting cluster halo \citep{Brough2008, Lidman2012, Lavoie2016, Kravtsov2018} and they are frequently found close to, and typically have relatively small velocity offsets with respect to, the potential centres of the cluster halo \citep{Lidman2013, Lauer2014}.  This is commonly interpreted as an indication that the formation and evolution of the BCGs and their host clusters are intimately linked, and that studying the former will provide clues about the formation and evolution of the latter.  However, pinning down the assembly and growth histories of these gigantic galaxies is proving to be a challenge.

There are several pathways that could explain the origin of the BCGs: (i) extended in-situ star formation (e.g. via cooling flows); (ii) rapid star formation and early assembly; and (iii) early star formation in separate galaxies but relatively recent assembly via a sequence of late-time mergers.  The key idea underlying the first proposal is that radiative cooling drives the hot intracluster medium to concentrate at the clusters' potential centre where it then forms stars at relatively high rates \citep{Cowie1977, Fabian1977}.  However, observations show that not only are the bulk of the BCG stars old \citep{Whiley2008}, but heating from the central active galactic nucleus (AGN) also strongly suppresses the cooling of the intracluster medium \citep{Tabor1993, Ciotti1997, Silk1998}.  The second proposal suggests that massive elliptical galaxies, including the BCGs, form via essentially monolithic collapse of a mass density peak \citep{Eggen1962}, with the galaxies' stellar mass building up rapidly in the process.  One difficulty with this model is that BCGs show evidence of significant growth in their sizes over cosmic time \citep{Daddi2005, VanderWel2008, Shankar2015}.  Once a leading theory, this scenario has fallen out of favour due to the emergence of the hierarchical assembly paradigm for cosmic structure formation.  According to this paradigm, the third proposal, galaxies form via a series of mergers of lower-mass systems -- implying that the low mass systems form first, and over time build-up the more-massive systems \citep{Aragon-Salamanca1998, Dubinski1998}.

Numerical studies investigating the formation and evolution of BCGs in the presently favoured hierarchical $\Lambda\mathrm{CDM}$ model find that the majority of stars that end up in the present-day BCGs typically form at $z \gtrsim 4$ in distinct progenitor galaxies \citep{DeLucia2006}.  These galaxies then eventually merge to assemble the BCGs we observe today.  As for the timing of this assembly, until recently, the theoretical consensus was that present-day BCGs are assembled through dissipationless mergers, with between $50 - 60\%$ of their present-day stellar mass being incorporated at relatively late times, from $z \sim 1.5$ to the present \citep{Dubinski1998, Conroy2007, DeLucia2007, Ruszkowski2009, Laporte2013, Contini2014, Laporte2015}.  Recent cosmological hydrodynamical simulations suggest a slightly modified scenario \citep{Martizzi2016, Ragone-Figueroa2018} and highlight the importance of distinguishing between the galaxy proper and its extended stellar envelope.  Focusing on the galaxy proper (i.e. the mass within $50$ $\kpc$ of the centre), \cite{Ragone-Figueroa2018} find that while half of the stars that end up in the BCG have formed by $z \sim 3.7$, the assembly of the BCG occurs over an extended time-span and half of the BCGs' stellar mass only falls into place typically by $z \sim 1.5$.   In detail, the stellar mass of the BCGs grows on average by a factor of $2.5$ between $z = 2$ and $z = 0$, with a median stellar mass growth factor over all BCGs of $3.5$.  Interestingly, \cite{Ragone-Figueroa2018} also find that \textit{in-situ} star formation is not entirely negligible; it accounts for nearly $25\%$ of the growth in stellar mass between $z = 2$ and $z = 1$.   These minor revisions aside, the central paradigm -- of early formation of the stellar mass and late-time assembly of the BCG -- remains intact.

Recent observations, however, suggest that the late-assembly picture may not be as concrete as once thought, and that early assembly may play a major role in the evolution of the BCGs.  Several BCGs have been discovered at $z \sim 1 - 1.5$ that have stellar masses comparable to the most-massive galaxies in the universe \citep{Collins2009}.  If these BCGs were to grow at the rates theoretically expected via late-time hierarchical assembly, they would greatly exceed the predicted masses of those theoretical models.  Additionally, the debate on the size evolution of BCGs is far from settled, as there is also evidence that the luminosities and sizes of the BCGs, as a population, do not evolve much past $z \sim 1$ \citep{Whiley2008, Stott2011},  suggesting that little growth through the hierarchical scenario is possible.  The absence of observed evolution past $z \sim 1$ implies that these massive galaxies must grow via a combination of \textit{in-situ} star formation and early assembly.  With respect to the former, \cite{Webb2015} analyse a set of BCGs in the \textit{Spitzer} Adaptation of the Red-Sequence Cluster Survey (SpARCS; see \citealt{Muzzin2009, Wilson2009}) and find that a large contribution to the overall growth of the BCGs must be due to \textit{in-situ} star formation based on the estimated star formation rates of hundreds of BCGs (in the range $0.8 < z < 1.8$), and also find an increasing star formation rate with increasing redshift.  As to the latter, there is growing evidence of highly over-dense protocluster cores (e.g. \citealt{Ishigaki2015, Miller2018, Jiang2018, Higuchi2019}; Wang et al., in prep.; also see \citealt{Ito2019}, and \citealt{Overzier2016} for a broad census review) at high redshifts ($z \gtrsim 4$).  As we demonstrate in Section~\ref{sec:methods}, protocluster cores with a high density of galaxies are the birthplaces of BCGs, and highly-over-dense systems should collapse rapidly in the $\Lambda\mathrm{CDM}$ theory -- casting into doubt whether the theoretical consensus of late assembly is valid for the entire population of BCGs.

In this paper, we investigate the above tension between the current theoretical picture and accumulation of observational results in order to gain insight into the evolution of the BCG population.  Specifically, we use a bespoke non-cosmological simulation based on the observed parameters of the SPT2349-56 protocluster \citep{Miller2018} to track its forward evolution.  Our interest lies in determining the future evolution of the protocluster core -- including the fate of the observed galaxies -- and the timescale of its evolution.  We then use the MultiDark Planck 2 Bolshoi simulation \citep{Riebe2013, Klypin2016} a large-volume non-baryonic simulation, to estimate the frequency of similar events in the universe.  While dark-matter simulations exist that provide ample resolution and population statistics (through their large volumes) for discovering over-dense protoclusters at high redshift, simulating the equivalent volumes in tandem with the hydrodynamical equations of motion and galactic-baryonic physical processes is, at present, not feasible due to computational constraints.  These constraints force us to study the forward evolution of SPT2349-56 in the bespoke simulation.  In Section~\ref{sec:methods} we describe our setup and initial conditions for the SPT2349-56 simulation.  In Section~\ref{sec:stellar_growth}, we discuss the assembly and growth of the system.  In Section~\ref{sec:cluster_evolution},  we analyse a large volume dark-matter-only simulation in order to determine how frequent such highly over-dense events may be.  Lastly, we synthesise our findings and present a revised paradigm for the formation and the evolution of the BCGs in Section \ref{sec:conclusions}.

\section{Methodology}
\label{sec:methods}

We start by constructing a bespoke simulation of the SPT2349-56 system in order to specifically study its forward evolution.  Observations indicate that the 14 galaxies that comprise the core of the protocluster are within a $130$ $\kpc$ (physical) projected region on the sky at a mean redshift of $z \approx 4.3$, and we show the observed physical properties of each of the 14 galaxies in leftmost two columns of Table~\ref{tbl:estimatedprops}.  The observed line-of-sight velocity distribution, $\Delta V_\mathrm{LOS}$, was found to approximate a Gaussian distribution with $\sigma_\mathrm{LOS} = 408 \, \mathrm{km} \, \mathrm{s}^{-1}$ \citep{Miller2018}.  The cold gas masses, $\Mgas$, were estimated by converting the estimated\footnote{\cite{Miller2018} measured the CO(4-3) line luminosity and then converted to a CO(1-0) luminosity by using the ratio of the line brightness temperatures, calculated from a sample of sub-millimetre galaxies with measurements of both lines.} CO(1-0) line luminosity to a gas mass using a conservative estimate of the $\alpha_\mathrm{CO}$ conversion factor, $\alpha_\mathrm{CO} = 0.8$ $\Msun$ / $(\mathrm{K} \, \mathrm{km} \, \mathrm{s}^{-1} \, \mathrm{pc}^2)$.

In order to simulate the forward evolution of the system, we use a modified version of \pkg{GIZMO}\footnote{\url{http://www.tapir.caltech.edu/~phopkins/Site/GIZMO.html}} \citep{Hopkins2015a}, a publicly available gravity plus hydrodynamics simulation program, that is equipped with an implementation of the mesh-free finite mass method \citep{Lanson2008a, Lanson2008b, Gaburov2011}.

\begin{table*}
\centering
 \caption{Observed and estimated physical properties of SPT2349-56 \citep{Miller2018}. We use only the observed line-of-sight offset velocities ($\Delta V_\mathrm{LOS}$) and cold gas masses ($\Mgas$) to estimate the remaining physical parameters in this table.  $\Mstar$ is the estimated galactic stellar mass, $\Cvir$ is the estimated NFW halo concentration, $\Mvir$ is the estimated virial mass of each halo, and $\Vvir$ is the estimated virial velocity of the galaxy's host halo.}
 \label{tbl:estimatedprops}
 \begin{tabular}{ccccccc}
  \hline
  Label & $\Delta V_\mathrm{LOS}$ (km s$^{-1}$) & $\Mgas$ ($10^{10}$ $\Msun$) & $\Mstar$ ($10^{10}$ $\Msun$) & $\Mvir$ ($10^{10}$ $\Msun$) & $\Cvir$ & $\Vvir$ (km s$^{-1}$) \\                            
  \hline
  \hline
  A &  -90  & 12.0 & 5.14  & 514  & 1.67 & 537      \\
  B &  -124 & 11.2 & 4.79  & 479  & 1.69 & 524      \\
  C &  603  & 6.7  & 2.87  & 287  & 1.81 & 442      \\
  D &  -33  & 8.4  & 3.6   & 360  & 1.75 & 477      \\
  E &  84   & 4.8  & 2.05  & 205  & 1.89 & 395      \\
  F &  395  & 3.4  & 1.46  & 146  & 1.97 & 353      \\
  G &  308  & 1.6  & 0.685 & 68.5 & 2.18 & 274      \\
  H &  -719 & 4.4  & 1.88  & 188  & 1.91 & 384      \\
  I &  310  & 2.2  & 0.942 & 94.2 & 2.09 & 305      \\
  J &  481  & 2.2  & 0.942 & 94.2 & 2.09 & 305      \\
  K &  631  & 3.1  & 1.33  & 133  & 2.00 & 342      \\
  L &  -379 & 3.3  & 1.41  & 141  & 1.98 & 348      \\
  M &  34   & 1.2  & 0.514 & 51.4 & 2.26 & 249      \\
  N &  90   & 1.0  & 0.428 & 42.8 & 2.31 & 234      \\
  \hline
 \end{tabular}
\end{table*}

We simulate the protocluster in isolation with vacuum boundary conditions.  Specifically, we simulate the initial coalescence phase of the separate systems (i.e. galaxies with their own dark matter halo, gas, and stellar components) that make up the SPT2349-56 protocluster core.  At this early phase, we do not expect the cluster-scale dark matter envelope to be in place.  The simulation is non-cosmological, and evolves the equations of motion for $1$ $\Gyr$.  Gas and star particles in the simulation have an initial mass of $M_\mathrm{gas} = M_\mathrm{*} = 10^6$ $\Msun$.  Gas properties are calculated using the cubic spline kernel with $32$ neighbouring particles in our simulations.  The dark matter particle mass is $M_\mathrm{dark} = 5\times 10^6$ $\Msun$.  Additionally, we seed each galaxy with a black hole of mass $M_\mathrm{BH} = 10^5$ $\Msun$.  We use adaptive gravitational softening for all gravitationally interacting particles \citep{Hopkins2018}, which requires minimum softening parameters.  Baryonic particles have a minimum softening of $\epsilon_\mathrm{b, min} = 50 \, \mathrm{pc}$ and, for dark matter, we use a minimum softening of $\epsilon_\mathrm{dark, min} = 200 \, \mathrm{pc}$.

\subsection{Initial conditions}
\label{sec:initialconditions}

We assume a \cite{PlanckXVI} cosmology throughout the following procedure, and generate initial conditions using the \pkg{MakeGalaxy} software \citep{Hernquist1993, Springel1999, Springel2000, Springel2005b}.  Each galaxy in our synthetic SPT2349-56 system consists of a dark matter halo, gas disc and stellar disc, with no stellar bulge or surrounding gaseous circumgalactic medium.  Given the high overdensity of the SPT2349-56 system, we expect that the large-scale background dark matter field of the surrounding region contributes much less to the mass budget within $130$ $\kpc$ compared to the individual systems.  Therefore, we do not include an extended dark-matter field as the dynamics within the $130$ $\kpc$ region should not be affected. \pkg{MakeGalaxy} employs the methods of \cite{Springel2005b} to create stable spiral galaxies, which we summarise here.  We model the dark matter distribution in each galaxy with a \cite{Hernquist1990} profile where the scalelength is related to the corresponding NFW concentration $\Cvir$ of the halo \citep{Navarro1997}.  We model the gas and stellar discs with exponentially declining surface densities with the scalelength $H$ related to the angular momentum (through the spin parameter $\lambda$) of the system.  We follow \cite{Robertson2006} and use $\lambda = 0.033$, which is the mode of the spin distribution from cosmological simulations \citep{Vitvitska2002}.  The combined gas+stellar disc is also under the condition that the total disc mass is a fixed fraction of the total mass of the system, i.e. $M_\mathrm{disc} = m_\mathrm{d}M_\mathrm{vir}$.  To ensure disc stability we choose $m_\mathrm{d} = 0.03$ for each disc as values between $0.03 \lesssim m_\mathrm{d} \lesssim 0.05$ lead to stable discs in the $\Lambda\mathrm{CDM}$ cosmology \citep{Mo1998}.  The vertical structure of the stellar disc is that of an isothermal sheet with a radially constant scaleheight $z_\mathrm{0}$ given as a free parameter proportional to the scalelength of the disc, which we assume to be $z_\mathrm{0} = 0.1 H$.  For the vertical structure in the gas disc, we set the scaleheight such that hydrostatic equilibrium is enforced.  We tested each galaxy in isolation to ensure that the system is physically and numerically stable, and find an average star formation rate of $\approx 75$ $\Msunyr$ over $1$ $\Gyr$, with a peak of $\approx 140$ $\Msunyr$, for the most-massive galaxy.

In order to setup each galaxy, we require estimates of their dark matter halo, stellar, and gas masses as well as the concentrations of the dark matter halos.  We base our following calculations on the observed total galactic cold gas mass, $\Mgas$, shown in Table~\ref{tbl:estimatedprops}.  We prepare and simulate three separate realisations of the SPT2349-56 system.

\subsubsection{Masses}
\label{sec:ic_masses}

We estimate the dark matter halo mass of each individual galaxy in the protocluster core by first assuming a reasonable gas fraction $\gasfrac$, computing the corresponding galactic stellar mass $\Mstar$, and then estimating the halo virial mass $\Mvir$ from $\Mstar$.  

We define the gas fraction as,

\begin{equation}
\label{eq:gasfraction}
\gasfrac \equiv \frac{\Mgas}{\Mgas + \Mstar}
\end{equation}

\noindent and, therefore, the stellar mass is

\begin{equation}
\label{eq:halomass}
\Mstar = \bigg(\frac{1}{\gasfrac} - 1\bigg) \Mgas.
\end{equation}

\noindent We assume all of our galaxies have the same gas fraction, $\gasfrac = 0.7$, estimated from the results in \cite{Narayanan2012} and \cite{Tadaki2019}.  This assumption is reasonable for high-redshift galaxies, where gas fractions $f_\mathrm{gas} > 0.4$ are routinely inferred, with a large spread above this value \citep{Carilli2010, Daddi2010, Tacconi2010, Tacconi2013}.  In fact, the median gas fraction of protocluster galaxies measured in \cite{Tadaki2019} is $f_\mathrm{gas} \approx0.77$, comparable to our value, with a spread from $f_\mathrm{gas} \approx 0.4$ to $f_\mathrm{gas} \approx 0.9$.  We discuss the impact of varying $f_\mathrm{gas}$ in Section~\ref{sec:stellar_growth}.

As for relating $\Mvir$ and $\Mstar$, \cite{Behroozi2013a} show that the stellar-to-halo mass fraction is $\Mstar / \Mvir \sim 0.01$, within a factor of two for a wide range of halos at $z > 4$ that eventually become clusters of mass $\sim 10^{15}$ $\Msun$, the predicted $z = 0$ mass of the SPT2349-56 system (see Fig. 2 of \citealt{Miller2018}).  We expect that altering the stellar mass to halo mass ratio by a factor of $\sim 2$ will change our results by the same factor.  We do note that the results in \cite{Behroozi2013a} apply to central galaxies; however,  we assume that the 14 galaxies in the SPT2349-56 are the centrals of their own halos before commencing to merge, coalesce, and form the protocluster core region.  Using these results we estimate $\Mstar \approx 0.428 \Mgas$ and  $\Mvir \approx 42.8 \Mgas$ to within a factor of two.   We show the results of our calculations in Table \ref{tbl:estimatedprops}. 

\subsubsection{Halo properties}
\label{sec:ic_haloproperties}

Having an estimate of the virial mass of each system allows us to calculate the halo concentrations and virial velocities.  Using the universal halo concentration model from \cite{Bullock2001}, the concentration parameter is

\begin{equation}
\label{eq:haloconcentration}
\Cvir(\Mvir, z) \approx 9 \bigg( \frac{\Mvir}{M_\mathrm{coll,0}} \bigg)^{-0.13} (1 + z)^{-1},
\end{equation}

\noindent where $M_\mathrm{coll,0}$ is the typical collapsing halo mass at $z = 0$, with $M_\mathrm{coll,0} = 1.18\times 10^{13} \, \Msun$.  \cite{Diemer2015} show that, for halo masses in our regime of interest ($\Mvir \lesssim 4.42\times10^{12} \, \Msun$), equation~(\ref{eq:haloconcentration}) is an excellent approximation to their more general models.  The virial velocity follows from the virial mass as,

\begin{equation}
\label{eq:virialvelocity}
\Vvir \approx (10 \mathrm{G} M_\mathrm{vir} H(z))^{1/3},
\end{equation}

\noindent where $H(z) = H_\mathrm{0} \sqrt{\Omegamnow(1 + z)^{3} + \Omegalambnow}$.  We show the results of equations~(\ref{eq:haloconcentration}) and (\ref{eq:virialvelocity}), applied to each halo, in Table~\ref{tbl:estimatedprops}.

\begin{figure}
    \centering
    \includegraphics[width=\columnwidth]{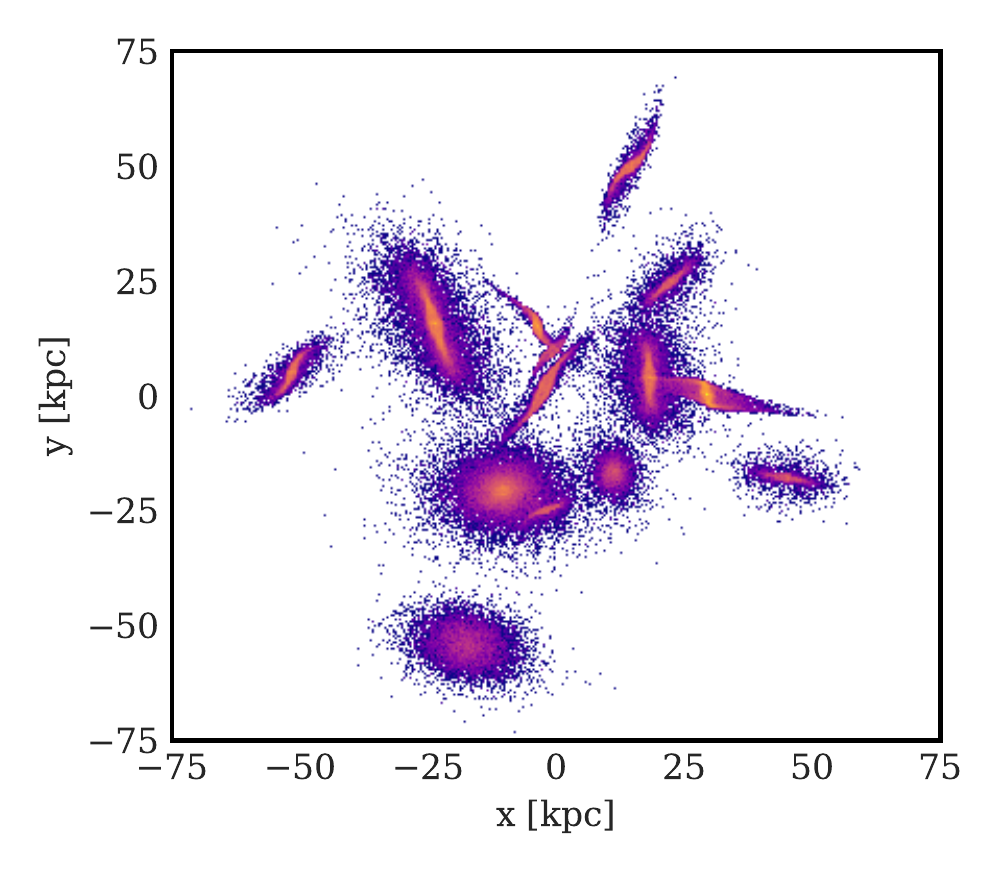}
    \caption{A schematic view of one of the realisations of our synthetic SPT2349-56 system.  We bin the \textit{xy}-plane positions of each stellar particle with an arbitrary logarithmic scaling at $1$ $\Myr$ after the initial condition.  We constrain the mass-weighted centres of each galaxy to lay within a sphere of radius $65$ $\kpc$ (physical) based on the observed projected separations.}
    \label{fig:initial_condition}
\end{figure}

\subsubsection{System dynamics}
\label{sec:ic_dynamics}

For the dynamical evolution of the entire system we require the initial positions and velocities of each galaxy.  We select initial positions randomly within a sphere of physical radius $65$ $\kpc$ (the observed maximal separation) for each galaxy with no dependence on the true separations between the observed objects.  Fig.~\ref{fig:initial_condition} shows a qualitative view of the positioning of each galaxy in one of the realisations of our synthetic SPT2349-56 system.  The observed velocity distribution in Table~\ref{tbl:estimatedprops} provides the initial velocities of our simulated galaxies.  We fit a Gaussian to the distribution and then sample 14 new velocity components for each spatial direction, assuming the velocity distribution is isotropic.  We randomly select the orientation of the spin axes for each galaxy.

\subsection{Galactic Physics}
\label{sec:methods_physics}

Our sub-grid physics models are the same as those described in \cite{Rennehan2019} (based on the model in \citealt{Dave2016c}), except that we now include a model for supermassive black hole (SMBH) growth and feedback.  We briefly describe the models below and point the reader to the aforementioned reference for more information.

\subsubsection{Cooling and Star Formation}
\label{sec:methods_coolsfr}

For radiative cooling, we calculate the cooling rates in the presence a UV background \citep{Faucher2009} using the \pkg{GRACKLE-3.1} cooling library\footnote{\url{https://grackle.readthedocs.io}} \citep{Smith2017}.

Our star formation implementation follows that in the MUFASA simulations \citep{Dave2016c, Dave2017}. We determine the conversion rate of gas into stars based on the estimated fraction of molecular hydrogen ($f_\mathrm{H_2}$) in the gas based on the approximations in \cite{Krumholz2009}.  We convert gas at densities above the threshold $n_\mathrm{crit} = 0.2 \, \mathrm{cm}^{-3}$ into stars at a rate $\mathrm{d}\rho_*/\mathrm{d}t = \epsilon_* f_\mathrm{H_2} \rho_\mathrm{gas} / t_\mathrm{dyn}$ where $\rho_*$ is the stellar density, $\epsilon_* = 0.02$ is star formation efficiency \citep{Kennicutt1998}, and $t_\mathrm{dyn} = (G\rho_\mathrm{gas})^{-1/2}$ is the local dynamical time.  We also force gas onto an artificial equation of state, $T_\mathrm{EoS} = 10^4 (n_\mathrm{gas} / n_\mathrm{crit})^{1/3} \, \mathrm{K}$, where $n_\mathrm{gas}$ is the gas hydrogen number density, above the star formation critical density ($n_\mathrm{gas} > n_\mathrm{crit}$) to suppress numerical fragmentation \citep{Teyssier2011, Dave2016c}.

\subsubsection{Stellar Feedback}
\label{sec:methods_feedback}

We include energetic feedback from supernova types Ia and II (SNIa and SNII, respectively), stellar radiation, and stellar winds from asymptotic giant branch (AGB) stars based on the MUFASA cosmological simulation model.  We also include mass injection from SNIa, SNII, and AGB stars, which is important for enriching the gas in the simulation \citep{Dave2016c, Liang2016}.  We account for the effects of both prompt and delayed SNIa \citep{Scannapieco2005}.

Metals are vital in determining the balance of gas cooling and heating in astrophysical gas, and therefore we include their production and account for their role in cooling.  We consider metal production by SNIa, SNII as well as AGB stars \citep{Iwamoto1999, Nomoto2006, Oppenheimer2008}.  For details, we refer the reader to \cite{Liang2016}, \cite{Dave2016c}, and \cite{Rennehan2019}.

\subsubsection{Active Galactic Nuclei}
\label{sec:methods_agn}

High luminosity galaxies often host active galactic nuclei (AGN) concurrently with intense starburst episodes in the local universe \citep{Nardini2008}, and at early epochs \citep{Alexander2005}.  Therefore, we also include the effects of AGN feedback into our investigation.  AGN are important in determining the correct estimate of stellar mass growth. Our model is that of \cite{Springel2005b}, which we briefly describe below.

We initially place black holes of mass $10^5$ $\Msun$ in the centres of each galaxy, and allow them to grow via Eddington-limited Bondi accretion.  We use the unboosted Bondi model because the mesh-free finite mass method can resolve higher densities at the same mass resolution, compared to the common smoothed particle hydrodynamics implementations \citep{Hopkins2015a}.  We also include energetic feedback, and assume that each AGN generates energy in the gas at a rate $\dot{E} = \epsilon_\mathrm{r} \epsilon_\mathrm{f} \dot{M}_\mathrm{BH} c^2$, where $\epsilon_\mathrm{r} = 0.1$ is the radiative efficiency, $\epsilon_\mathrm{f} = 0.05$ is the coupling fraction to the gas, and $c$ is the speed of light.  The accretion rate $\dot{M}_\mathrm{BH} = 4\pi G^2 M_\mathrm{BH}^2 \rho_\mathrm{gas} / (c_\mathrm{s}^2 + v_\mathrm{rel}^2)^{3/2}$ is the Bondi accretion rate onto the black hole, where $c_\mathrm{s}$ is the surrounding gas sound speed, $v_\mathrm{rel}$ is the relative velocity of the SMBH with respect to the gas, and $\rho_\mathrm{gas}$ is the surrounding gas density.  The gas properties -- density, sound speed, and relative velocity -- are calculated over the nearest $128$ neighbouring gas particles.  The energy is deposited to the surrounding gas in a kernel-weighted manner, over the same nearest $128$ neighbouring particles.  To follow the dynamical evolution of the SMBHs, we use the model from \cite{Tremmel2015a} in which the dynamical friction force is calculated by using the approximation from \cite{Chandrasekhar1943a}.

\section{Stellar assembly and growth}
\label{sec:stellar_growth}

\begin{figure}
    \centering
    \includegraphics[width=\columnwidth]{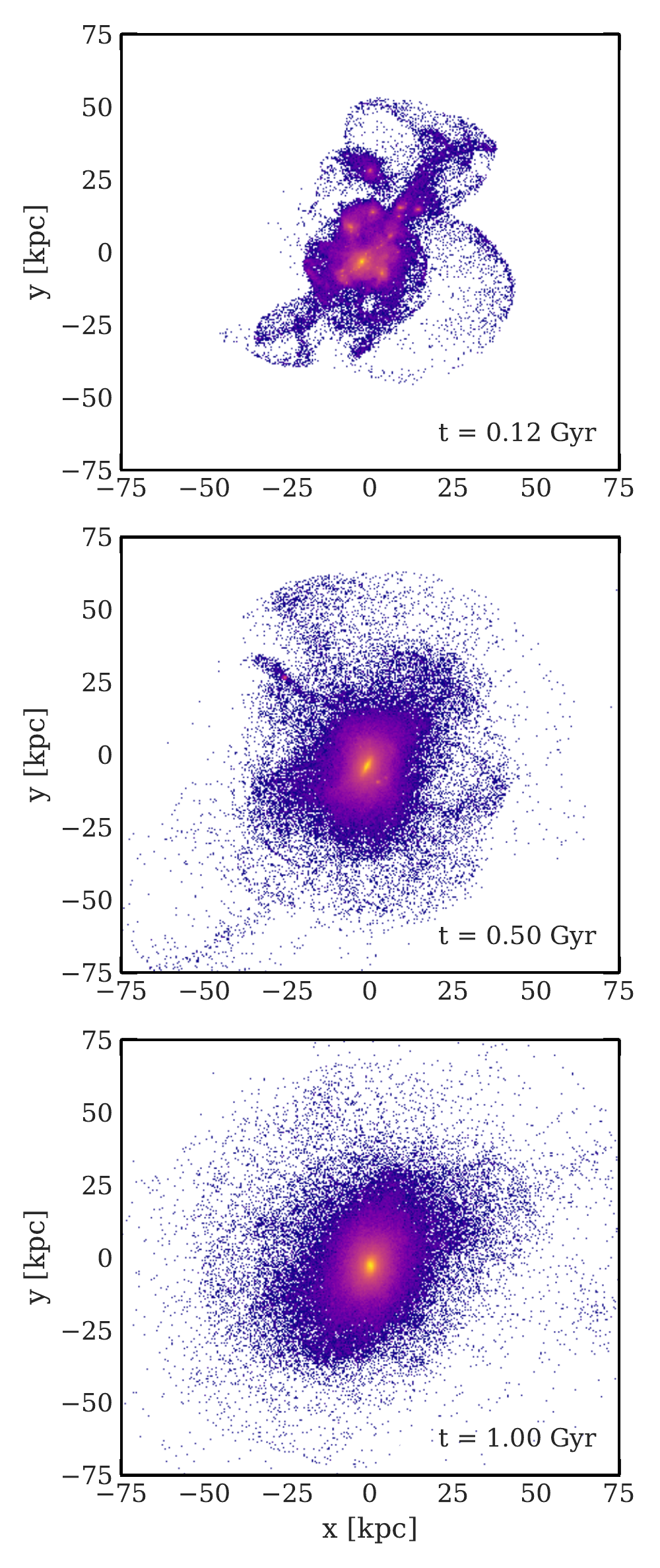}
    \caption{A particle view of the assembly of one of the realisations of our synthetic SPT2349-56.  We bin the positions of the stellar particles in the simulation in the \textit{xy}-plane ($150$ $\kpc$ $\times$ $150$ $\kpc$) using an arbitrary logarithmic scaling. These panels represent $t = 0.12$, $0.5$, and $1$ $\Gyr$ from the initial condition, from top to bottom, respectively.  There are obvious shell-like structures and stellar streams throughout the short assembly period, until the system resembles a massive elliptical galaxy at $\sim 500$ $\Myr$.}
    \label{fig:particle_plot}
\end{figure}

To gain a qualitative understanding of the protocluster assembly, we examine one realisation of the system visually in Fig.~\ref{fig:particle_plot}.  We bin the positions of each star particle in the simulated \textit{xy}-plane at three times: $t = 0.12$, $0.5$, and $1$ $\Gyr$, from top to bottom, respectively.

In the top panel of Fig.~\ref{fig:particle_plot}, there are several stellar streams protruding through the system as the galaxies undergo the initial collapse after $\sim 120$ $\Myr$.  These streams are due to tidal stripping from the companion galaxies as the initial velocity dispersion of the system, combined with the close proximity of galaxies, is unable to prevent imminent merging.  At maximum distance, the tidal tails extend approximately $90$ $\kpc$.

In the middle panel, $500$ $\Myr$ after the start of the simulation, several streams are visible in addition to shell-like structures surrounding the core of the galaxy.  It is difficult to distinguish any of the original structure as the stellar populations begin to mix.  At this point during the simulation, many of the stellar particles are launched out to $\approx 75$ $\kpc$ from the centre-of-mass of the system, building up the diffuse stellar envelope and intracluster light.

At $1$ $\Gyr$ in the bottom panel of Fig.~\ref{fig:particle_plot}, the system resembles a massive elliptical galaxy.  Many of the stars from the initial galaxies, as well as those formed \textit{in-situ}, were kicked out and formed an extended diffuse stellar halo.  There is no longer any visible structure in this halo.  Using radiative transfer we are able to further investigate the evolution of the system.

\begin{figure*}
    \centering
    \includegraphics[width=0.8\textwidth]{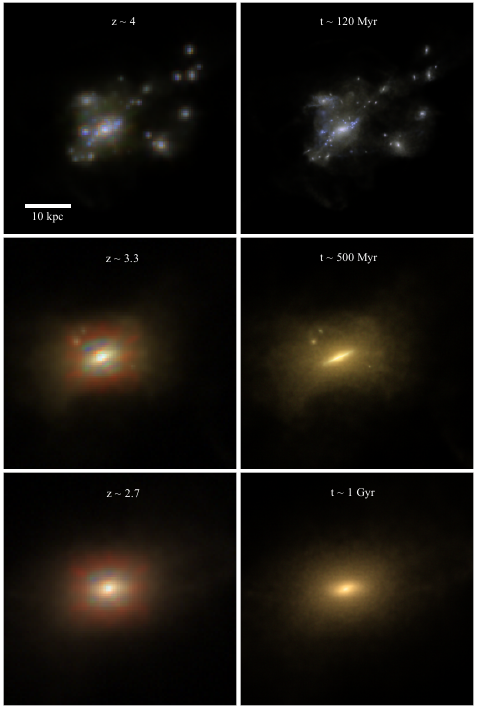}
    \caption{(\textit{left}) James Webb Space Telescope mock observations of the forward evolution of SPT2349-56.  Redshifts are $z \sim 4$, $3.3$, and $2.7$, from top to bottom, respectively. (\textit{right}) Synthetic false colour images made from three of the James Webb Space Telescope NIRCam filters without the CCD angular scale or noise.  The snapshots correspond to the redshift labels in the left column, for each row.  The scale marker shows $10$ $\kpc$ (physical) and each panel is $50$ $\kpc$ per side.}
    \label{fig:jwst_mock_observations}
\end{figure*}

Fig.~\ref{fig:jwst_mock_observations} shows colour-composite mock James Webb Space Telescope (JWST) observations of one of our synthetic SPT2349-56 realisations.  We generated these observations using the \pkg{SKIRT} radiative transfer code \citep{Baes2011, Camps2015, Baes2015} and an adapted version of the observational realism suite described in \cite{Bottrell2017a, Bottrell2017b}.  To produce these images, we first assigned star particles spectral templates based on the STARBURST99 \citep{Leitherer1999} spectral energy distribution set for old stellar populations and \citet{Groves2008} templates, which include emission from \textsc{HII} and photodissociation regions, for young ($<10$ Myr-old) star particles.  We use a multi-component dust model \citep{Zubko2004} with a constant dust-to-metal ratio of $0.3$ and do not limit dust to star-forming gas particles.  \pkg{SKIRT} produced rest-frame optical data cubes that we processed into noiseless, idealised photometric images in the JWST NIRCam {F150W, F200W, F277W, F356W} band passes\footnote{\url{https://jwst-docs.stsci.edu/near-infrared-camera/nircam-instrumentation/nircam-filters}} at redshifts corresponding to the simulation snapshots.  In the right column of Fig.~\ref{fig:jwst_mock_observations}, we show idealised images.  The idealised images are noiseless and are neither rebinned down to the NIRCam angular resolution nor convolved with the NIRCam point-spread function.  At each of the redshifts we consider $z \in \{4, 3.3, 2.7\}$, corresponding to $t \in \{0.12, 0.5, 1\}$ $\Gyr$, at least three of these filters reside in the protocluster's redshifted rest frame optical domain.  

We assigned the filtered light at the smallest wavelengths to the blue values, the mid-range wavelengths to the green, and the longest to the red to construct a qualitative, visual representation of the system from $t \sim 120$ $\Myr$ to $t \sim 1$ $\Gyr$ (from top to bottom in Fig.~\ref{fig:jwst_mock_observations}, respectively)\footnote{Specifically, we use the NIRCam filters \{F200W, F277W, F356W\} at $120$ $\Myr$, and \{F150W, F200W, F277W\} at both $500$ $\Myr$ and $1$ $\Gyr$.}. 

In the top-right panel of Fig.~\ref{fig:jwst_mock_observations}, we show $t \sim 120$ $\Myr$ after the initial condition.  Already by this point in the system's evolution, we see the long stellar streams from the top panel of Fig.~\ref{fig:particle_plot} as a spatially-extended low-surface brightness web surrounding the remnant cores of each initial galaxy.  Several bright star-forming cores remain visible in blue and white.  Initial gravitational torques cause the fragmentation we observe in the protocluster early on due to the close proximity of the 14 galaxies.

At $500$ $\Myr$ in the centre row on the right, there are a few remnant cores of the original galaxies and the system assumes a more spherical shape in the diffuse stellar halo.  By this point in time, the gas settles into an extended disc in the centre of the gravitational potential, and we see the system shift toward a redder appearance.  We analysed the simulation data and discovered that the disc only exists from $t \sim 300$ $\Myr$ to $t \sim 600$ $\Myr$, before the remaining cores of the original galaxies dynamically pummel the disc, leading to its demise.  It is not surprising that there is a disc given the high gas fraction of the system \citep{Hopkins2009}.

At $1$ $\Gyr$, the system is elliptical and resembles a typical low-redshift cD galaxy.  Although the disc vanishes, star formation remains at an absolute rate of $\sim40$ $\Msunyr$, or a specific rate of $5\times10^{-11}$ $\mathrm{yr}^{-1}$, much lower than values predicted in \cite{Webb2015} for BCGs at slightly lower redshift ($z \sim 2$).

In addition to the idealised mock observations in the right column of Fig.~\ref{fig:jwst_mock_observations}, we show mock observations of the protocluster in the left column.  We rebin the idealised images to the NIRCam CCD angular scale corresponding to the appropriate channel ($0.031$ $\mathrm{arcsec} \, \mathrm{pixel}^{-1}$ for F200W and $0.063$  $\mathrm{arcsec} \, \mathrm{pixel}^{-1}$ for the others), point-spread function convolution, and added noise corresponding to the predicted NIRCam surface-brightness sensitivity for our observing strategy.   Using the simulated ramp-optimization feature of the JWST \pkg{pynrc} package (Leisenring et al., "pyNRC: A NIRCam ETC and Simulation Toolset", in prep.), we determined the optimal observing strategy for the protocluster assuming $10 \, \mathrm{ks}$ of observing time.  For more details, see Appendix~\ref{app:mock_observations}.  We analyse these results at three selected redshifts: $z \in \{4, 3.3, 2.7\}$, from top to bottom in Fig.~\ref{fig:jwst_mock_observations}, respectively.  These leftmost panels correspond to the same redshift as the idealised images in the same row in Fig.~\ref{fig:jwst_mock_observations}.

In the top-left panel, the substructure at $z \approx 4$ is clearly visible in the JWST composite image.  We find that the absolute AB-magnitude in the F277W NIRCam band is $M_{\mathrm{AB},\mathrm{F277W}} \approx -28.7$, making our synthetic SPT2349-56 an extremely bright object.  In an analysis of 430 brightest cluster galaxies in \cite{Donzelli2011}, the brightest galaxy (object A0401 in the Table 2 of that study) has an absolute \textit{R}-band magnitude of $-27.25$.  Although our filters are not equivalent, we find the same approximate magnitude order in all of the JWST filters we apply and, therefore, we can conclude that SPT2349-56 would end up being one of the brightest galaxies in the observable universe.  Not only is our realisation bright, it is also very blue.  We calculated the rest-frame SDSS $\mathrm{g - r}$ colours for the system and find $\mathrm{g - r} \approx -0.05$ at $z \approx 4$ for our dust model.  A typical BCG has a redder colour at $\mathrm{g - r} \approx 0.6$ \citep{Cerulo2019}.

The qualitative view of the BCG at a redshift of $z \approx 3.3$ corresponds to $500$ $\Myr$.  At this time, the stars remain blue in the system ($\mathrm{g - r} \approx 0.13$), which is interesting since we do not expect fully assembled, highly star-forming BCGs at $z \approx 3.3$ with a large fraction of young stellar populations.  If the predictions of substantial (i.e. factors of $\sim 2$ to $\sim 4$) growth past $z \sim 1$ in the literature \citep{DeLucia2007, Lavoie2016, Ragone-Figueroa2018} are correct, then systems such as SPT2349-56 would go on to become the most-massive galaxies in the universe at $z = 0$, reaching stellar masses up to $\sim 3 \times 10^{12}$ $\Msun$.  We do not include any of the substructure that may exist outside of the core of the SPT2349-56 protocluster, which could increase our estimate even further if those galaxies were to accrete between $z \sim 3$ and $z \sim 1$ via dynamical friction (i.e. galactic cannibalism; \citealt{Ostriker1975, White1976}).  However, since they grow rapidly to a large mass, the dynamical friction timescale for lower-mass satellites to merge into the system becomes large, which could hamper late-time growth. 

\begin{figure}
	\centering
	\includegraphics[width=\columnwidth]{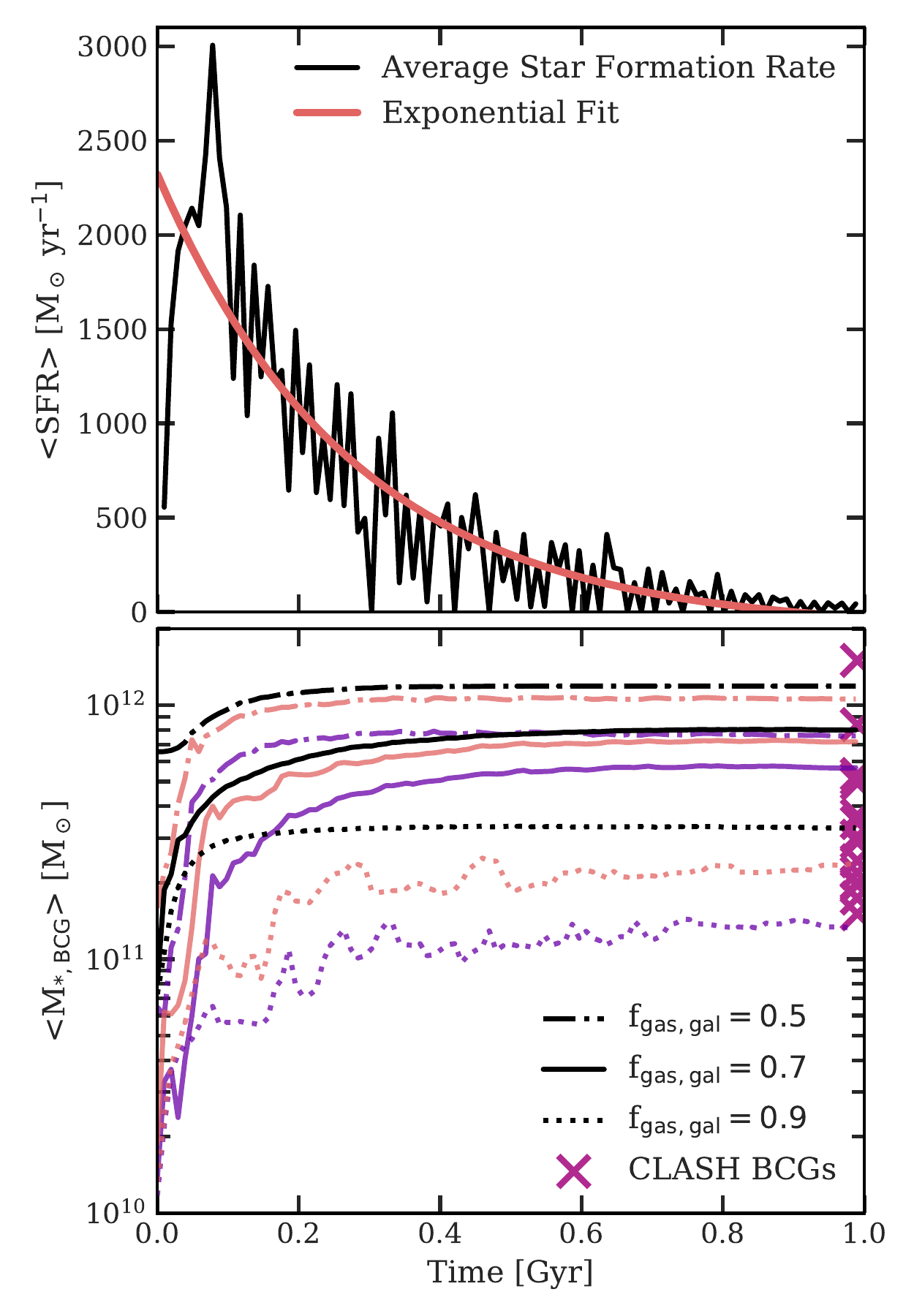}
	\caption{Growth and assembly history of the protocluster simulations. (\textit{top}) Star formation rate over the first giga-year of evolution averaged over three independent realisations of our fiducial simulation with initial gas fraction $\gasfrac = 0.7$.  We used a non-linear least squares method to fit a decaying exponential and found a decay timescale of $\tau \approx 200$ $\Myr$. (\textit{bottom}) The black curves show the total mass of all stellar particles in the simulation volume for each system with different initial gas fraction.  The purple and coral curves show the stellar mass within $5$ $\kpc$ and $15$ $\kpc$ apertures, centred on the peak projected stellar mass, respectively.    We show the observed masses of brightest cluster galaxies from the CLASH survey as $\times$ symbols at $1$ $\Gyr$ which were derived using \pkg{MAG\_AUTO} magnitudes.  After $1$ $\Gyr$ ($z \approx 2.7$), the total stellar masses across the three simulations with varying gas fractions is larger than most of the lower-redshift CLASH brightest cluster galaxies.}
	\label{fig:proto_sfr}
\end{figure}

In the top panel in Fig.~\ref{fig:proto_sfr}, we show the forward-evolution of the star formation rate of our synthetic SPT2349-56 system (i.e. our fiducial simulation with $\gasfrac = 0.7$) as a function of time, averaged over three realisations, up to $1$ $\Gyr$ ($z \approx 2.7$) after the observation ($z \approx 4.3$)\footnote{In terms of redshift, our simulation began at $z \approx 4.3$ and ended $1$ $\Gyr$ later at $z \approx 2.7$, assuming a \cite{PlanckXVI} cosmology.}.   The star formation rate (SFR) peaks at $\sim 3000$ $\Msunyr$ approximately $5$ $\Myr$ after the initial condition and decays exponentially to $\sim 40$ $\Msunyr$ at $1$ $\Gyr$.  Our simulated SFR is comparable to the observed results of \cite{Miller2018} who find a total star formation rate of $\approx 6500$  $\Msunyr$ for the $14$ galaxies in the SPT2349-56 system.  We use a non-linear least squares fitting method to fit an exponential curve to the SFR, and find that the decay time is $\tau \sim 200$ $\Myr$.  Initially, the SFR peaks due to high compression from the proximity of each individual galaxy, and the fact that they are rapidly collapsing under their mutual gravity.  Additionally, strong fluctuations in SFR occur as the discs undergo tidal interactions.  Stellar feedback is the main cause of the declining SFR, as the black holes provide little feedback past the initial $\sim 50$ $\Myr$ due to their low accretion rates.  They are unable to accrete because of their high velocities relative to the medium -- which suppresses accretion.  On the other hand, stellar feedback is most powerful when the star formation rate is the highest, as the system coalesces, and the supernovae and stellar winds begin to excavate the gas out of the original galaxies.

The balance of stellar feedback and star formation causes a smooth decline in star formation rate, and consequently a smooth increase in the stellar mass, as we show in the bottom panel in Fig.~\ref{fig:proto_sfr}.  The total stellar mass of our fiducial simulation is given as the solid black line with initial and final (at $1$ $\Gyr$) stellar masses, $M_\mathrm{initial} = 2.81\times 10^{11}$ $\Msun$ and $M_\mathrm{final} = 8.14\times 10^{11}$ $\Msun$, respectively. The total stellar mass is the sum of all of the individual stellar particle masses in the entire simulation volume.  The assembly mass, which we define as 90\% of the total final mass, is $M_\mathrm{assembly} = 7.33\times 10^{11}$ $\Msun$.  The time at which assembly occurs is $t_\mathrm{assembly} \sim 370$ $\Myr$.  Assembly occurs rapidly in the system, and our results show that the system nearly quadruples size within a giga-year.

Apart from the total stellar mass, we also examine the growth of the stellar mass of largest galaxy in the simulation, which we label as the evolving protocluster’s BCG.  Observationally, the stellar masses of BCGs are usually determined within an aperture with a specified radius since only the projected stellar light distribution is known.  Therefore, to compare to observational stellar masses, it is more reasonable that we choose an aperture centred on the peak stellar brightness and sum the stellar mass along the corresponding cylinder.  \cite{Kravtsov2018} advocate apertures of fixed radii (they used $30$ $\kpc$, $50$ $\kpc$, and $70$ $\kpc$) for comparing simulated BCG masses to observed masses.  \cite{Burke2015}, whose masses we compare to below, adopted aperture sizes for each BCG in the Cluster Lensing and Supernova survey with Hubble (CLASH) \citep{Postman2012} sample such that it encompassed $\sim90\%$ of the flux.  \cite{Stott2010} used the same method to determine the BCG stellar masses for the systems in the XMM Cluster Survey (XCS; \citealt{Romer2001, Sahlen2009}), and showed that a fixed $50$ $\kpc$ aperture gave similar results.  A $50$ $\kpc$ aperture is, however, too large for our high-redshift protocluster since the diameter of the system is only $135$ $\kpc$.  A $50$ $\kpc$ aperture centered on the peak in the projected stellar mass distribution\footnote{We tested multiple viewing angles and found that our results do not change significantly.} -- approximating the location of the peak surface brightness -- encompasses $95\%$ of the total stellar mass (black curves in Fig.~\ref{fig:proto_sfr}) over course of the simulation.  To isolate the most massive galaxy in our simulated protocluster and track the growth of its stellar mass over time, we use smaller apertures of $5$ $\kpc$ and $15$ $\kpc$ since the effective radius (i.e. the half-light radius) of the stellar distribution is $\approx 2.2$ $\kpc$ at $1$ $\Gyr$ in our fiducial simulation and, therefore, the apertures enclose a few effective radii.

The purple and coral solid curves in the bottom panel of Fig.~\ref{fig:proto_sfr} show the stellar mass growth within the aforementioned cylindrical apertures of radii $5$ $\kpc$ and $15$ $\kpc$ apertures, respectively.  The smaller apertures more closely track the mass growth in the most massive galaxy across the entire simulation although the merging process is so rapid that distinguishing between the BCG and the debris from in-falling, merging, disrupting galaxies past $\sim 100$ $\Myr$ is impractical, and the differences between the stellar masses within each aperture become less significant.  The masses with the apertures rise rapidly as the galaxies coalesce over the course of the first $\sim100$ $\Myr$ (see the top panel of Fig.~\ref{fig:particle_plot}) and converge to their final values by $200 - 250$ $\Myr$.  At $1$ $\Gyr$, the $5$ $\kpc$ and $15$ $\kpc$ apertures contain $\approx 70\%$ and $\approx 90\%$ of the total stellar mass, respectively.

We also show the observed masses of brightest cluster galaxies (BCGs) in \cite{Burke2015} from the CLASH survey with $\times$ symbols at $1$ $\Gyr$.  These BCGs belong to clusters with masses $\gtrsim 10^{15}$ $\Msun$, and range in redshift from $0.187 < z < 0.890$.  Although our system is at $z \sim 2.7$ after $1$ $\Gyr$, the total stellar mass and the mass within the smaller apertures are already more massive than most of the BCGs from the CLASH survey.  We emphasise that we only mean to compare the observed BCG masses with our simulated BCG at $1$ $\Gyr$, after the system resembles a fully-formed BCG (see the bottom panel of Fig.~\ref{fig:particle_plot}).

It is pertinent that we emphasise that we designed our numerical experiment to be simple in nature to determine the star formation rate and assembly timescale of the protocluster.  We did not tune the star formation rate to match the observations and, given that  we are within a factor of $\sim2$, the broad assumptions appear reasonable.  The main sources of error are (i) our choice for the gas fraction and (ii) the abundance matching results.  To test the dependence on gas fraction, we simulated two additional realisations with a lower ($\gasfrac = 0.5$) and higher ($\gasfrac = 0.9$) gas fraction.  We show the total stellar mass evolution in the protocluster system for each initial gas fraction in the bottom panel of Fig.~\ref{fig:proto_sfr}.

In the case of $\gasfrac = 0.5$, the total initial stellar mass and, therefore, dark matter mass are a factor of $\sim 2.3$ larger than in our fiducial simulation.  This is because we fix our gas masses based on the observations in \cite{Miller2018}.  We find that the total stellar mass of this low gas fraction system at $1$ $\Gyr$ is $\sim1.2\times 10^{12}$ $\Msun$, which is a factor of $\sim 1.5$ higher than in our fiducial simulation but still much more massive than the majority of CLASH BCGs.  The dashed purple and coral curves in Fig.~\ref{fig:proto_sfr} show that the growth within $5$ $\kpc$ and $15$ $\kpc$ is rapid in this system and the most massive galaxy quickly grows in size.  At $1$ $\Gyr$, $\approx 99\%$ of the total stellar mass is contained within the $5$ $\kpc$ aperture and $\approx 90\%$ is contained within the $15$ $\kpc$ aperture.  Based on our definition of the stellar assembly time, we find $t_\mathrm{assembly} \sim 150$ $\Myr$ which is a factor of $\sim 2.5$ smaller than in the fiducial simulation.  Our low gas fraction system therefore reaches a slightly higher stellar mass and does so more quickly. 

Next, we consider the simulation with a high gas fraction of $\gasfrac = 0.9$.  This results in an initial stellar mass a factor of $\sim 4$ lower than our fiducial simulation and, therefore, the same factor lower in halo mass.  The total stellar mass at $1$ $\Gyr$ of the simulated system is $\sim3.3\times 10^{11}$ $\Msun$ -- a factor of $\approx 2.5$ lower than our fiducial simulation.  This BCG is much less concentrated than the other BCGs with lower gas fractions as we find $\approx 50\%$ and $\approx 85\%$ of the total stellar mass in the apertures of radii $5$ $\kpc$ and $15$ $\kpc$, respectively.  We find that the assembly time is a factor of $\sim 3$ shorter than in our fiducial simulation at $t_\mathrm{assembly} \sim 120$ $\Myr$.

In summary, the above variations in the initial gas fractions result in only factors of $2$--$3$ in the final stellar mass and assembly time.    Therefore, we argue that our results are relatively robust; we expect that all scenarios involving factors of $\sim 2$ changes in $\gasfrac$ or $\Mstar / \Mvir$ will result in a massive BCG at high-redshift that assembles its stellar mass very quickly (within $\sim 1$ $\Gyr$).  We attribute the differences in assembly timescales between the simulations with different initial gas fractions as due to the varying potential well depth.  In the lowest gas fraction case, the potential is deeper and the interactions are stronger, therefore the assembly process occurs on shorter timescales.  In the high gas fraction case, the potential well is shallower and stellar feedback, following the initial intense star-burst, leads to the expulsion of gas -- quenching the system more rapidly than in the fiducial case.

As for the choice of $\Mvir \sim 0.01 \Mstar$ (see Fig. 8 of \citealt{Behroozi2013a}), for halos at $z > 4$ which become halos of mass $\sim 10^{15}$ $\Msun$ at $z = 0$, the stellar mass to halo mass ratio is $\sim0.01$ to within a factor of two.  Decreasing (increasing) the ratio would lower (raise) the virial mass of the halo, which in turn will alter the dynamics of the system at the level of a factor $\sim2 - 3$, similar to what we discussed above.

As an additional source of uncertainty, \cite{Granato2015} point out that it is difficult to reproduce high star formation rates in high-redshift protoclusters in numerical simulations using standard sub-grid models of star formation, stellar feedback, and active galactic nuclei feedback.  However, the sub-grid models we used in \pkg{GIZMO} have been shown to reproduce broad galaxy population properties \citep{Dave2017, Dave2019} at a wide-range of redshifts and in high-redshift ($z \sim 6$) galaxies at similar particle mass resolutions \citep{Olsen2017}.

We posit that recently discovered systems similar to SPT2349-56 (e.g. \citealt{Ishigaki2015, Jiang2018, Higuchi2019}) are the proto-cores of the massive galaxy clusters.  The speed of assembly and growth of stellar mass in our simulated realisations of SPT2349-56 is obvious from Figs.~\ref{fig:particle_plot}, ~\ref{fig:jwst_mock_observations}, \&~\ref{fig:proto_sfr}.  The star formation rate declines exponentially with an \textit{e}-folding time of $\sim200$ $\Myr$ while simultaneously, an object that qualitatively looks like many observed BCGs forms by $500$ $\Myr$.  From these results, we predict that the observed high-redshift over-dense protoclusters (\citealt{Ishigaki2015, Miller2018, Jiang2018}; Wang et al., in prep.) will undergo a similar evolution and therefore form massive BCGs as early as $z \approx 4$.  As we demonstrated, the JWST will be able to clearly see the massive BCGs out to redshift of $z \approx 3$, and their progenitors out to $z \approx 4.3$, opening up a new frontier for exploration, especially in collaboration with survey telescopes such as the Wide-Field Infrared Survey Telescope (WFIRST). 

\section{Implications for galaxy clusters}
\label{sec:cluster_evolution}

In the hierarchical structure formation scenario, galaxy clusters are the youngest and most-massive objects in the universe.  In this theory, the more massive the cluster, the younger and rarer the system.  Hence, the most-massive clusters should continue to assemble the bulk of their mass until late times.  What we have demonstrated in the previous section is the possibility of a \textit{downsizing} effect \citep{Bower2006, Cimatti2006, Neistein2006, Fontanot2010, Oser2010} on the cluster scale, where the cores of these massive clusters could be much older than the cores of less-massive clusters, beginning to assemble at redshifts $z \gtrsim 2$ at the minimum.  Our results on the rapid assembly of the brightest cluster galaxy (BCG) are then not unexpected as these high-redshift protocluster cores would be the strongest relative overdensities in the early universe.  This begs the question: why has the theoretical community not predicted high-redshift fully-assembled BCGs?

One of the issues is that the main tools of contemporary theoretical astrophysicists are numerical simulations.  These are necessary because of the strong non-linearity of the structure formation process after the linear perturbation theory breaks down.  However, a lack of computational power limits the spatio-temporal dynamic range of the numerical simulations.  We are therefore forced into a compromise.  We can model a small comoving volume of the universe and resolve the galaxies in this volume (e.g. \citealt{Schaye2014, Pillepich2018, Dave2019}) but such volumes generally do not contain the rare massive clusters we would expect to host an object such as SPT2349-56.  Or, we can sacrifice resolution in favour of large volumes but then galaxy formation must be introduced in an ad hoc fashion \citep{Ruszkowski2009}, which can introduce biases.  Despite these difficulties, we do expect that these protoclusters are present in the largest dark-matter-only simulations, such as the MultiDark Bolshoi \citep{Klypin2011} and Millennium XXL \citep{Angulo2012} simulations.

To test our theory of early, rapid assembly of the cores of the most-massive clusters with regards to the general population of galaxy clusters, we shift focus from SPT2349-56 and now investigate the assembly history of the most massive clusters (at $z = 0$) in the MultiDark Planck 2 (MDPL2) simulation \citep{Riebe2013, Klypin2016} -- a child of the MultiDark Bolshoi suite of simulations.  Our goal is to determine if there is a population of BCGs that assemble early and their relationship to their host clusters.  The MDPL2 simulation consists of $3840^3$ particles within a simulation volume of side-length $1$ $\mathrm{c}\Gpch$, and has a mass resolution of $1.51\times 10^9$ $\Msunh$.  All of the halo data is publicly available online\footnote{\url{https://www.cosmosim.org}} and we specifically use the \pkg{MDPL2.Rockstar} database in the following analysis.  This database contains halo properties that were determined using the \pkg{ROCKSTAR} halo finder \citep{Behroozi2013}, and includes the substructure trees for each host halo.

First, we require a set of criteria for the occurrence of highly over-dense massive collapse events at high redshift. We analysed the formation histories of all galaxy clusters that had final masses\footnote{Henceforth we assume a \cite{PlanckXVI} cosmology for $h$ in our quoted masses and distances.}  of $\Mvir$ $ \geqslant 5\times 10^{14}$ $\Msun$ at $z = 0$, and narrowed our search to those with a large number of relatively massive halos entering their progenitor's virial radius across cosmic time.  Specifically, we define an over-dense collapse event to be when $N \geqslant 5$ halos of individual mass $\Mvir$ $ \geqslant 2\times 10^{11}$ $\Msun$ all enter the virial radius of a more massive halo within a time $\Delta t$, where $\Delta t$ is the time between two consecutive simulation snapshots.  For example, at $z \approx 4.5$ this timescale corresponds to $\sim 40$ $\Myr$ in the \pkg{MDPL2.Rockstar} database.  We also ensure that there are at least 5 lower mass halos that are no more than $20$ times less massive than the more massive halo.  The latter condition considers only those events in which the substructures have a chance of merging within a Hubble time at their respective redshift due to dynamical friction, estimated at the virial radius of the more massive halo \citep{Mo2010},

\begin{equation}
    t_\mathrm{df}|_{r_\mathrm{i}=R_\mathrm{vir}} \approx \frac{1.17}{\ln(M_\mathrm{large} / M_\mathrm{sub})}\bigg(\frac{M_\mathrm{large}}{M_\mathrm{sub}}\bigg)\frac{1}{10H(z)},
    \label{eq:dynamical_friction}
\end{equation}

\noindent where $M_\mathrm{large}$ is the virial mass of the more massive system, $M_\mathrm{sub}$ is the mass of the substructure, and $H(z) = H_\mathrm{0} \sqrt{\Omegamnow(1 + z)^{3} + \Omegalambnow}$.  Each structure is within a factor of $\sim20$ of the more massive halo and, therefore, $t_\mathrm{df}(z) \sim 3 t_\mathrm{H}(z) / 4$ where $t_\mathrm{H}(z) = 1 / H(z)$ is the Hubble time.  We consider the dynamical friction timescale to be a conservative estimate of how quickly the systems will merge since our criteria captures all mergers with $M_\mathrm{large} / M_\mathrm{sub} \lesssim 20$ and the merging halos are not on circular orbits at the virial radius; therefore, they should merge faster \citep{Poole2006, Boylan-Kolchin2008}.

\begin{figure}
    \centering
    \includegraphics[width=\columnwidth]{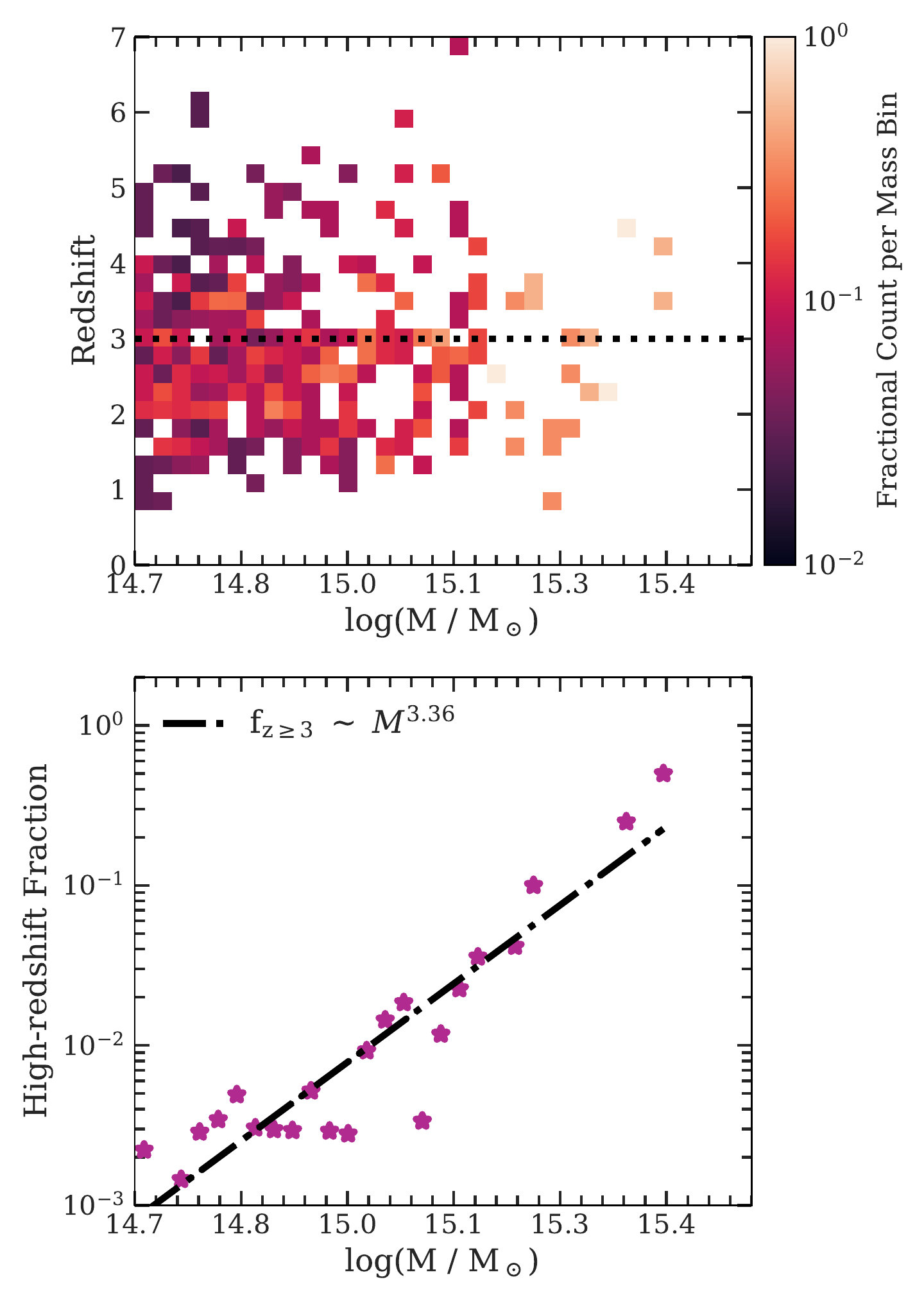}
    \caption{(\textit{top}) A two-dimensional histogram showing the probability of the first occurrence of a highly over-dense region collapsing at a specific redshift, in a given bin of final cluster virial mass (i.e., the virial mass at $z = 0$) spanning from $5\times 10^{14}$ $\Msun$ to $3.25\times 10^{15}$ $\Msun$ (the most-massive cluster).  We remove non-events from this histogram so that the fraction is truly the probability of finding an event at a specific redshift, given that such an event occurs. There is a slight trend for more instances of the events at higher final cluster masses but the scatter in the distributions shows that many events occur for all of the clusters here above $z = 3$.  (\textit{bottom}) The number of over-dense collapse events above $z = 3$, normalised to the total number of clusters in each mass bin.  The bins are identical to those in the top panel.  The dash-dot line shows a power-law fit with $f \sim M^{3.36}$ -- high-redshift over-dense collapse events are more frequent as a function of increasing final cluster mass. }
    \label{fig:event_frequency} 
\end{figure}

We also assume that the substructures and the host each contain a galaxy with stellar mass given by the abundance matching relations in \cite{Behroozi2013a}.  One important caveat to note is that the results of \cite{Behroozi2013a} apply to central galaxies and not satellites.  However, since the substructures are independent halos before merging, this implies that our criteria is equivalent to a minimum of $6$ central galaxies merging.  For comparison, the most massive galaxy in our scenario plays a similar role to galaxy A in Table~\ref{tbl:estimatedprops} and in the analysis of Section~\ref{sec:stellar_growth}.  

We determine the first redshift at which an over-dense event occurs for each of the galaxy clusters above our mass limit.  The least massive events we find are a factor of $\sim20$ less massive than our estimated mass for SPT2349-56 and ought to rapidly (if found above $z \gtrsim 3$) collapse into elliptical galaxies of mass $M_\mathrm{*} \gtrsim 10^{11}$ $\Msun$, if we assume the same abundance matching estimates from Section~\ref{sec:methods} combined with the aforementioned dynamical friction timescale constraint. 

In the top panel of Fig.~\ref{fig:event_frequency}, we present a two-dimensional histogram that encodes the occurrence of high-density collapse regions for a cluster of a given final mass at $z = 0$.  Specifically, we count the number of first occurrences of an over-dense event in a grid of redshift and final cluster mass coordinates.  We normalise to the total number of events in each mass bin, so that the colouring shows the probability of finding the first event for that final cluster mass at a given redshift, compared to all other events that occur.  We do not include non-events in the top panel as our interest lies in the probability of an event occurring at a specific redshift, given all of the events that occur.  

From the top-panel in Fig.~\ref{fig:event_frequency}, lower-mass clusters at $z = 0$ are less likely to have a high-density collapse at redshifts $z \gtrsim 3$ (above the dashed line in Fig.~\ref{fig:event_frequency}) compared to clusters with masses $\Mvir$ $ \gtrsim 10^{15}$ $\Msun$.  There are, however, clusters that did not experience an over-dense collapse at any redshift.

The trend of increasing event frequency with increasing final cluster mass is more evident in the bottom panel of Fig.~\ref{fig:event_frequency}, where we show the number of events above $z \geqslant 3$ in each mass bin, normalised to the total number of clusters in each mass bin.  We show a fit to the fraction as a function of $z = 0$ cluster mass, and find that the fraction scales as $f \sim M^{3.36}$.  There is a clear power-law trend where the fraction of events increases with increasing $z = 0$ mass of the clusters.  In the highest mass bin, there is a $\approx 50\%$ chance of all clusters having a high-redshift over-dense collapse event occur at $z \geqslant 3$.

For the present purposes, our interest lies in the objects that collapse in the range $z \gtrsim 3$ since these would end up as the most-massive, blue, elliptical galaxies before $z \sim 1.5$ by our predictions.  We find that $16.4\%$ of clusters with final masses $\Mvir$ $ > 10^{15}$ $\Msun$ have an over-dense collapse event occur above $z = 3$, and $9.8\%$ of clusters with final masses $5\times 10^{14}$ $\Msun$ $\leqslant $ $\Mvir$ $ < 10^{15}$ $\Msun$.  In total, we find $155$ high-density collapse events at $z \geqslant 3$.  Given the volume of the simulation, $V \approx 3.2$ $\mathrm{c}\Gpc^{3}$, we expect a comoving number density of corresponding massive, blue, elliptical cluster BCGs at redshifts $z \gtrsim 1.5$ to be $n \approx 48$ $\mathrm{c}\Gpc^{-3}$ (or $n \approx 4.8\times 10^{-8}$ $\mathrm{c}\Mpc^{-3}$).

\begin{figure}
    \centering
    \includegraphics[width=\columnwidth]{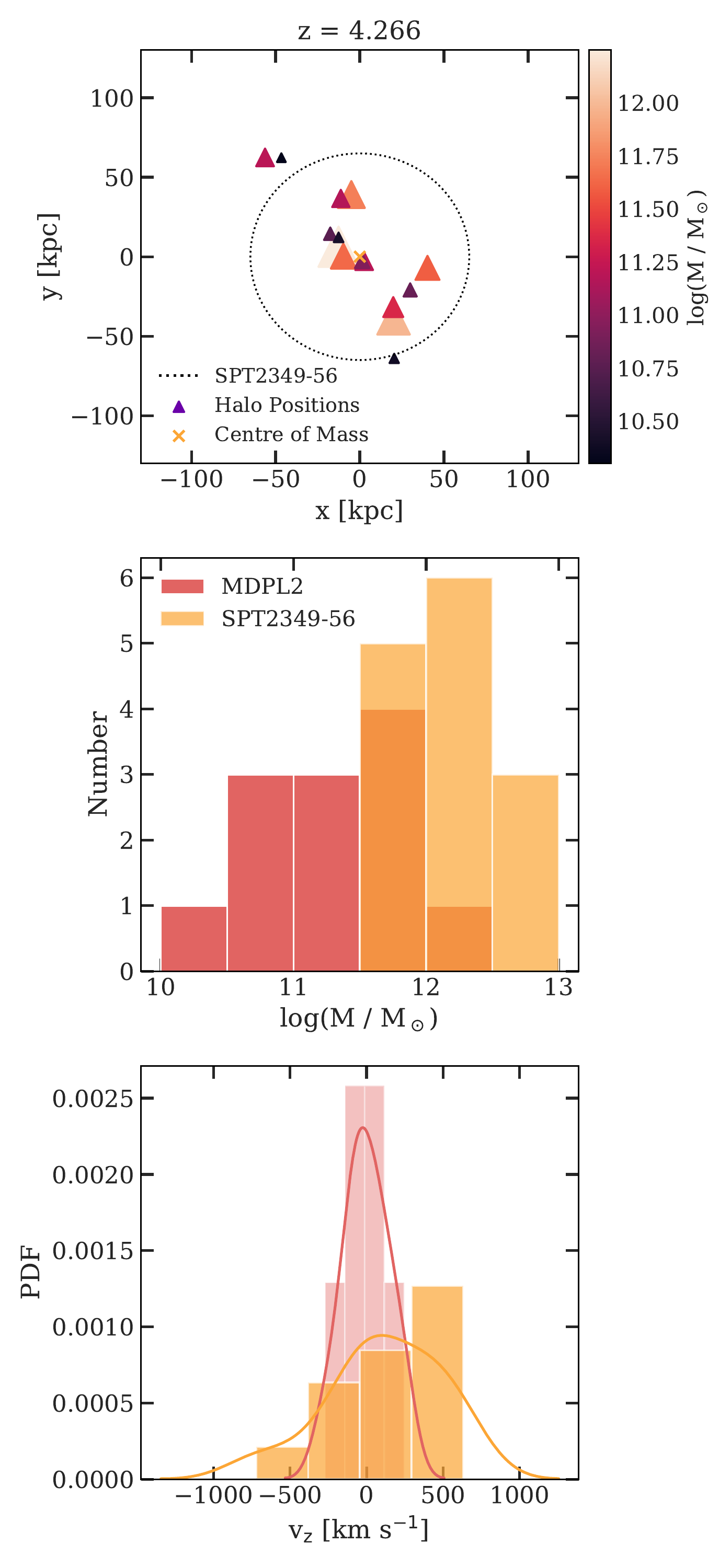}
    \caption{An example of one of the many high-redshift, over-dense regions we found in the Multi-Dark Planck 2 simulation.  This system is the progenitor of the second-most-massive galaxy cluster at $z = 0$, and is approximately an order of magnitude less in mass than our estimation of SPT2349-56. (\textit{top}) Spatial (physical) positions of substructure within a more-massive halo ($\Mvir$ $\sim 10^{13}$ $\Msun$).  The physical projection of the observed SPT2349-56 region is overlaid with a dotted line. (\textit{middle}) Virial mass histogram of substructure compared with the estimated virial masses of the dark matter halos in SPT2349-56. We consider only those masses that lay within the dotted line.  (\textit{bottom}) Kernel density estimate of the substructure velocities along the line of sight.  We show the SPT2349-56 observed values as a comparison.  As above, we only consider substructure within the dotted circle in the top panel for the velocity distribution.}
    \label{fig:protocluster_example}
\end{figure}

In Fig.~\ref{fig:protocluster_example}, we examine the spatial, mass, and velocity distribution (from top to bottom, respectively) of an example cluster proto-core in the MDPL2 simulation\footnote{Specifically, we examine the halo with \pkg{rockstarId}=$923730455$ in the \pkg{MDPL2.Rockstar} table.}.  This protocluster is the most-massive progenitor (MMP) at $z = 4.266$ (the approximate redshift of SPT2349-56) of the second-most-massive cluster at $z = 0$.

In the top panel of Fig.~\ref{fig:protocluster_example}, we show the spatial distribution of the substructure entering the virial region of the MMP.  The dotted line shows the observed extent of the SPT2349-56 object for comparison.  Each halo is marked with a triangle that we colour based on the logarithmic mass of the halo, and we arbitrarily scale the size of each triangle to the virial radius of each halo.  Furthermore, we mark the centre of mass of the system with a $\times$ symbol.  The virial radius of the MMP is $R_\mathrm{vir,phys} \approx 120$ $\kpc$ and we see that most of the massive substructure is within the SPT2349-56 sphere, or approximately half of the virial radius of the MMP.  In this particular system, there are eight halos with masses $M_\mathrm{sub} > 10^{11}$ $\Msun$, and we assume that they will merge on timescales shorter than the dynamical friction timescale taken at the virial radius of the host following equation~\ref{eq:dynamical_friction}.  At $z = 4.266$ the Hubble time is $t_\mathrm{H} \approx 1.43$ $\Gyr$ assuming a \cite{PlanckXVI} cosmology and, therefore, $t_\mathrm{df} \approx 1$ $\Gyr$.  Of course, this is a reasonable estimate since the substructure is within half of the virial radius of the more massive system and, therefore, will merge much more quickly.  Evidently, over-dense systems such as the example in the top panel assemble rapidly in the early universe.  However, will they go on to form objects as massive and bright as our synthetic SPT2349-56? 

To address this question, we show the mass distribution of our cluster proto-core in the middle panel of Fig.~\ref{fig:protocluster_example}.  As we mention above, there are eight objects above $M_\mathrm{sub} > 10^{11}$ $\Msun$, and we first assume that each of these halos host a galaxy.  Using the same methodology as in Section~\ref{sec:ic_masses}, we estimate that each halo should host a galaxy with stellar mass that is $1\%$ of the halo mass.  Therefore, each galaxy in the system should contain, on average, a stellar mass of $M_* \sim 5\times10^{10}$ $\Msun$.  If we use a conservative estimate of mass doubling within $1$ $\Gyr$ of evolution based on Section~\ref{sec:stellar_growth}, we would expect this object to become an elliptical galaxy with stellar mass $M_* \sim 10^{11}$ $\Msun$ at $z \approx 2.7$ and, like the BCG in our simulation, it would be extremely bright and blue at that redshift.

In the bottom panel of Fig.~\ref{fig:protocluster_example} we show a comparison of the velocity differences between SPT2349-56 and our example system.  We calculate the velocity offset in the perpendicular direction to the top panel, relative to the centre of mass of the system, in order to mimic how \cite{Miller2018} determined the velocities of their object\footnote{The velocity distribution of the simulated halos does not change appreciably with respect to the projection plane.}.  The solid lines show the kernel density estimate of the underlying distribution for both our system and SPT2349-56.  The velocity dispersion of our example system is $\sigma \approx 590$ $\mathrm{km}\, \mathrm{s}^{-1}$ and we find a tightly peaked distribution within the extent of the velocity dispersion, indicating that the system is indeed bound, with no signs of filamentary structure within the virial radius.  In SPT2349-56, we see that there is a bias toward approaching velocities, indicating that the system is most likely being fed by a filament. 

Our investigation into the MDPL2 simulation suggests that there is a non-negligible fraction ($\sim10\%$) of BCGs that assemble their stars extremely rapidly in the early universe.  This counters the notion of late-time assembly for the entire population of BCGs, and especially counters the idea of treating the entire set of BCGs as a homogeneous population to begin with.   What we have not addressed is the overall trend of later assembly with decreasing cluster mass and what this means for the BCG population.  Using the same aforementioned timescale estimates, systems that we identify as highly over-dense regions that collapse at $z \sim 3$ should go on to form BCGs with stellar masses $M_* \gtrsim 10^{11}$ $\Msun$ by $z \sim 1.5$.  Therefore, our results suggest that there is a continuous population of bright, star-forming BCGs down to $z \sim 1.5$ -- where the Hubble time is still short, allowing for the rapid assembly of substructure.  Recent observations have indeed begun to undercover this population of star-forming galaxies \citep{McDonald2016}, especially with \cite{Webb2015} finding that the star formation rate in their BCG sample increases to $\sim 1000 - 3000$ $\Msunyr$ as a function of increasing redshift.  Similarly, the core of the recently investigated galaxy cluster SpARCS104922.6+564032.5 at $z = 1.71$ \citep{Webb2017, Trudeau2019} appears qualitatively similar to our results from Section~\ref{sec:stellar_growth}, and the scenario we propose above could explain the extended star formation and morphology within that system.  

There is an additional connection of these high-redshift BCGs to galaxy clusters -- one that could help ameliorate discrepancies in the determination of cosmological parameters.  Specifically, the value of the density fluctuation power spectrum amplitude, $\sigmaeight$, derived from galaxy cluster counts is known to disagree at $\sim 2\sigma$ with the value derived from the cosmic microwave background in the Planck mission \citep{Ade2014, Douspis2018, Salvati2018}.   The number of massive clusters is highly sensitive to $\sigmaeight$ and we expect these BCGs to preferentially exist in those clusters.  In this study, through our use of the MDPL2 simulation, we use the \cite{PlanckXVI} value of $\sigmaeight$ which is higher than those derived from cluster counts.  As it is, events such as SPT2349-56 are rare in this cosmology and ought to be rarer still if $\sigmaeight$ were even slightly lower.  A combination of surveying with the Wide-Field Infrared Survey Telescope (WFIRST) and follow-up confirmation with the JWST would provide a lower-bound on the number density, hence providing an additional constraint on $\sigmaeight$ that could settle the issue.  While we provide only an approximate estimate of the number density of massive BCGs, a more detailed study could provide an exact constraint on the value.

One major caveat is that for the BCGs to be highly star-forming, they must not exhaust their gas supply, and they must not have their star formation halted by active galactic nuclei feedback.  However, given that the regions we discovered at $z \gtrsim 3$ are highly over-dense, they undergo what could be considered an effectively \textit{monolithic} formation scenario where the substructure gives rise to galaxies that are gas rich, which then rapidly merge to form the BCG.  And only afterwards, after the newly formed BCG's gas content depletes due to feedback and consumption, would star formation quench and the galaxy age passively.

\section{Conclusions}
\label{sec:conclusions}

The cores of galaxy clusters are home to the most luminous galaxies in the universe -- the brightest cluster galaxies (BCGs).  These galaxies have unique properties, such as their velocity dispersion and luminosity profiles, that set them apart from other galaxies at the high-end of the galaxy luminosity function.  The contemporary picture of their formation and growth scenario is that their stars are old and formed at high redshift ($z \gtrsim 4$) in separate individual galaxies that, at late times ($z \lesssim 1$), hierarchically assemble to form the massive galaxies we observe today.  There are, however, recent observations of highly-overdense protoclusters at $z \gtrsim 4$ \citep{Ishigaki2015, Miller2018, Jiang2018} that muddle this simple picture, since we expect these to go on to form the BCGs \citep{Ito2019}.

The hierarchical assembly picture of structure formation predicts the gradual build-up of objects through successive mergers of small objects.  Under this theory, we expect the most-massive galaxies to be the youngest in terms of their assembly time.  However, observations of massive ellipticals show a \textit{downsizing} effect whereby the mass of the systems and the stellar ages are anti-correlated \citep{Cimatti2006, Bower2006, Fontanot2010}.  In other words, the most-massive systems appear to have the oldest stellar populations.  We propose a new paradigm wherein a similar downsizing effect occurs on the scale of galaxy clusters themselves. In our proposal, the cores of the most-massive clusters -- the BCGs -- assemble earlier than the cores of lower-mass clusters, on average.  A subset of cluster-cores are assembled at very high redshift ($z \gtrsim 3$),  with the probability of high-redshift assembly decreasing as a function of decreasing mass (at $z = 0$) of the clusters.  

In order to determine the rapidity of assembly and growth of the BCGs, we studied the forward-evolution of a recently discovered, highly over-dense protocluster at $z \approx 4.3$, SPT2349-56 \citep{Miller2018} using a non-cosmological hydrodynamic simulation informed by the observations.  We found that the star formation peaks at $\sim 3000$ $\Msunyr$ very early in our simulation, and decays exponentially with a timescale of $\tau \sim 200$ $\Myr$.  By $1$ $\Gyr$, we found that the system remains at a stable star formation rate of $\sim 40$ $\Msunyr$.  We found that the system assembles $90\%$ of its mass at $t_\mathrm{assembly} \sim 370$ $\Myr$ after the initial condition and has a SDSS rest-frame $\mathrm{g - r}$ colour of $\mathrm{g - r} \approx 0.13$.  In terms of redshift, $370$ $\Myr$ corresponds to $z \sim 3.3$ and implies that fully formed, highly star-forming, blue BCGs should exist at redshifts $z \gtrsim 3.3$, given that there are observations of systems similar to SPT2349-56 above $z \gtrsim 4.3$ \citep{Ishigaki2015, Jiang2018}.  We demonstrated that new observational tools such as the James Webb Space Telescope will be able to easily image such systems, given that we estimated their absolute magnitudes to be $M_{\mathrm{AB},\mathrm{F277W}} \sim -28.7$ in the F277W NIRCam band.  Of course, these estimations depend on the assumptions we made when constructing our numerical experiment.  The main sources of uncertainty are the gas fractions and the applicability of the abundance matching results \citep{Behroozi2013} to the $z \approx 4.3$ protocluster.  However, all high-redshift galaxies are found with relatively high gas fractions and while we are near the boundary of applicability in the abundance matching results, we are still within the halo mass range where the results apply.

We expect that systems such as SPT2349-56 go on to form the cores of the most-massive clusters in the universe.  We used the Multi-Dark Planck 2 simulation\footnote{A $(1$ $\mathrm{c}\Gpch)^3$ volume with $3840^3$ dark matter elements.} -- a child of the Multi-Dark Bolshoi simulations -- to investigate the occurrence of highly over-dense assembly events in the early universe for all clusters with $z = 0$ masses $\Mvir$ $ \geqslant 5\times 10^{14}$ $\Msun$.  Our analysis revealed that there is a clear trend of less-massive clusters having over-dense events occur at lower redshifts (compared to more-massive clusters), indicating that there is a \textit{downsizing} effect.  Specifically, we found that the fraction of events above $z \geqslant 3$ for a given bin of final cluster mass scales strongly with the final cluster mass, $f \sim M^{3.36}$.  We also determined that $\sim10\%$ of all the clusters we investigated begin to assemble their cores rapidly at high redshift, with a higher percentage of $16.4\%$ for clusters with final masses above $\Mvir$ $ \geqslant 10^{15}$ $\Msun$.  Based on these estimates, we predict that there is a population of bright, blue BCGs above $z \gtrsim 1.5$ with a comoving number density of $n \approx 48$ $\mathrm{c}\Gpc^{-3}$.  Additionally, we predict that there is a similar star-forming BCG population that extends down to $z \sim 1.5$ -- although those at lower redshift will be up to an order-of-magnitude less massive than their high-redshift counterparts, given the downsizing trend.  At redshifts lower than $z \sim 1.5$, there is insufficient gas to support high star formation rates and those BCGs that begin assembling late will assemble through dry mergers, in keeping with the conventional picture.  We emphasise the distinction between core assembly and assembly of the rest of the cluster, and that our results do not suggest that the entire cluster assembles at high-redshift.  We expect protocluster cores to be embedded in an extended lower density (compared to the proto-core) galaxy distribution, with these galaxies eventually forming the satellite population of the assembling cluster.

Given that extraordinary infrared observational tools such as Wide-Field Infrared Space Telescope (WFIRST) and the JWST will launch in the upcoming decade,  we expect the census of interesting astronomical objects to broaden significantly.  Based on our arguments in this paper, we anticipate that some of those high-redshift objects will be the cores of the most-massive clusters in the universe -- the BCGs. Not only are these interesting objects in their own right but their discovery and census would also help to constrain the discrepancy in the measurements of the cosmological parameter $\sigmaeight$, given their rarity and association with the most-massive clusters.

\section*{Acknowledgements}

This research was enabled in part by support provided by WestGrid and Compute/Calcul Canada.  Some of the computations in this work were performed on facilities supported by the Scientific Computing Core at the Flatiron Institute. The Flatiron Institute is supported by the Simons Foundation.  DR acknowledges the support of the Natural Sciences and Engineering Research Council of Canada (NSERC), [funding reference number 534263].  DR and AB acknowledge support from NSERC (Canada) through the Discovery Grant program.  CB acknowledges the support of the Natural Sciences and Engineering Research Council of Canada (NSERC), [funding reference number 504189].  MHH acknowledges the receipt of a Vanier Canada Graduate Scholarship.  We thank Chervin Laporte, Carlos Frenk, Julie Hlavacek-Larrondo, Laura Salvati, Axel Weiss, Mark Brodwin, Gabriella De Lucia, Michael McDonald, Belaid Moa, and Ondrea Clarkson for insightful discussions during the course of the study.  We also thank the anonymous reviewer for their helpful comments which helped us improve our study.

The CosmoSim database used in this paper is a service by the Leibniz-Institute for Astrophysics Potsdam (AIP).  The MultiDark database was developed in cooperation with the Spanish MultiDark Consolider Project CSD2009-00064.  The authors gratefully acknowledge the Gauss Centre for Supercomputing e.V. (\url{www.gauss-centre.eu}) and the Partnership for Advanced Supercomputing in Europe (PRACE, \url{www.prace-ri.eu}) for funding the MultiDark simulation project by providing computing time on the GCS Supercomputer SuperMUC at Leibniz Supercomputing Centre (LRZ, \url{www.lrz.de}).  The Bolshoi simulations have been performed within the Bolshoi project of the University of California High-Performance AstroComputing Center (UC-HiPACC) and were run at the NASA Ames Research Center.

Our analysis was performed using the Python programming language (Python Software Foundation, \url{https://www.python.org}).  The following packages were used throughout the analysis: \pkg{h5py} \citep{Collette2013}, \pkg{numpy} \citep{Oliphant2006, VanderWalt2011}, and \pkg{matplotlib} \citep{Hunter2007}.  Part of the analysis was performed in the IPython environment \citep{Perez2007}.  This research additionally made use of \pkg{astropy} (\url{http://www.astropy.org}) -- a community-developed core Python package for Astronomy \citep{astropy2013, astropy2018}.




\bibliographystyle{mnras}
\bibliography{SPT2349} 

\begin{thebibliography}{}
\makeatletter
\relax
\def\mn@urlcharsother{\let\do\@makeother \do\$\do\&\do\#\do\^\do\_\do\%\do\~}
\def\mn@doi{\begingroup\mn@urlcharsother \@ifnextchar [ {\mn@doi@}
  {\mn@doi@[]}}
\def\mn@doi@[#1]#2{\def\@tempa{#1}\ifx\@tempa\@empty \href
  {http://dx.doi.org/#2} {doi:#2}\else \href {http://dx.doi.org/#2} {#1}\fi
  \endgroup}
\def\mn@eprint#1#2{\mn@eprint@#1:#2::\@nil}
\def\mn@eprint@arXiv#1{\href {http://arxiv.org/abs/#1} {{\tt arXiv:#1}}}
\def\mn@eprint@dblp#1{\href {http://dblp.uni-trier.de/rec/bibtex/#1.xml}
  {dblp:#1}}
\def\mn@eprint@#1:#2:#3:#4\@nil{\def\@tempa {#1}\def\@tempb {#2}\def\@tempc
  {#3}\ifx \@tempc \@empty \let \@tempc \@tempb \let \@tempb \@tempa \fi \ifx
  \@tempb \@empty \def\@tempb {arXiv}\fi \@ifundefined
  {mn@eprint@\@tempb}{\@tempb:\@tempc}{\expandafter \expandafter \csname
  mn@eprint@\@tempb\endcsname \expandafter{\@tempc}}}

\bibitem[\protect\citeauthoryear{Alexander, Smail, Bauer, Chapman, Blain,
  Brandt  \& Ivison}{Alexander et~al.}{2005}]{Alexander2005}
Alexander D.~M.,  Smail I.,  Bauer F.~E.,  Chapman S.~C.,  Blain A.~W.,  Brandt
  W.~N.,   Ivison R.~J.,  2005, \mn@doi [Nature] {10.1038/nature03473}, 434,
  738

\bibitem[\protect\citeauthoryear{Angulo, Springel, White, Jenkins, Baugh  \&
  Frenk}{Angulo et~al.}{2012}]{Angulo2012}
Angulo R.~E.,  Springel V.,  White S. D.~M.,  Jenkins A.,  Baugh C.~M.,   Frenk
  C.~S.,  2012, \mn@doi [Monthly Notices of the Royal Astronomical Society]
  {10.1111/j.1365-2966.2012.21830.x}, 426, 2046

\bibitem[\protect\citeauthoryear{Aragon-Salamanca, Baugh  \&
  Kauffmann}{Aragon-Salamanca et~al.}{1998}]{Aragon-Salamanca1998}
Aragon-Salamanca A.,  Baugh C.~M.,   Kauffmann G.,  1998, \mn@doi [Monthly
  Notices of the Royal Astronomical Society]
  {10.1046/j.1365-8711.1998.01495.x}, 297, 427

\bibitem[\protect\citeauthoryear{{Astropy Collaboration}}{{Astropy
  Collaboration}}{2013}]{astropy2013}
{Astropy Collaboration} 2013, \mn@doi [Astronomy {\&} Astrophysics]
  {10.1051/0004-6361/201322068}, 558, A33

\bibitem[\protect\citeauthoryear{{Astropy Collaboration}}{{Astropy
  Collaboration}}{2018}]{astropy2018}
{Astropy Collaboration} 2018, \mn@doi [The Astronomical Journal]
  {10.3847/1538-3881/aabc4f}, 156, 123

\bibitem[\protect\citeauthoryear{Baes \& Camps}{Baes \& Camps}{2015}]{Baes2015}
Baes M.,  Camps P.,  2015, \mn@doi [Astronomy and Computing]
  {10.1016/j.ascom.2015.05.006}, 12, 33

\bibitem[\protect\citeauthoryear{Baes, Verstappen, {De Looze}, Fritz, Saftly,
  {Vidal P{\'{e}}rez}, Stalevski  \& Valcke}{Baes et~al.}{2011}]{Baes2011}
Baes M.,  Verstappen J.,  {De Looze} I.,  Fritz J.,  Saftly W.,  {Vidal
  P{\'{e}}rez} E.,  Stalevski M.,   Valcke S.,  2011, \mn@doi [The
  Astrophysical Journal Supplement Series] {10.1088/0067-0049/196/2/22}, 196,
  22

\bibitem[\protect\citeauthoryear{Behroozi, Wechsler  \& Wu}{Behroozi
  et~al.}{2013a}]{Behroozi2013}
Behroozi P.~S.,  Wechsler R.~H.,   Wu H.-Y.,  2013a, \mn@doi [The Astrophysical
  Journal] {10.1088/0004-637X/762/2/109}, 762, 109

\bibitem[\protect\citeauthoryear{Behroozi, Wechsler  \& Conroy}{Behroozi
  et~al.}{2013b}]{Behroozi2013a}
Behroozi P.~S.,  Wechsler R.~H.,   Conroy C.,  2013b, \mn@doi [The
  Astrophysical Journal] {10.1088/0004-637X/770/1/57}, 770, 57

\bibitem[\protect\citeauthoryear{Bottrell, Torrey, Simard  \& Ellison}{Bottrell
  et~al.}{2017a}]{Bottrell2017a}
Bottrell C.,  Torrey P.,  Simard L.,   Ellison S.~L.,  2017a, \mn@doi [Monthly
  Notices of the Royal Astronomical Society] {10.1093/mnras/stx017}, 49, 1033

\bibitem[\protect\citeauthoryear{Bottrell, Torrey, Simard  \& Ellison}{Bottrell
  et~al.}{2017b}]{Bottrell2017b}
Bottrell C.,  Torrey P.,  Simard L.,   Ellison S.~L.,  2017b, \mn@doi [Monthly
  Notices of the Royal Astronomical Society] {10.1093/mnras/stx276}, 467, 2879

\bibitem[\protect\citeauthoryear{Bower, Benson, Malbon, Helly, Frenk, Baugh,
  Cole  \& Lacey}{Bower et~al.}{2006}]{Bower2006}
Bower R.~G.,  Benson A.~J.,  Malbon R.,  Helly J.~C.,  Frenk C.~S.,  Baugh
  C.~M.,  Cole S.,   Lacey C.~G.,  2006, \mn@doi [Monthly Notices of the Royal
  Astronomical Society] {10.1111/j.1365-2966.2006.10519.x}, 370, 645

\bibitem[\protect\citeauthoryear{Boylan-Kolchin, Ma  \&
  Quataert}{Boylan-Kolchin et~al.}{2008}]{Boylan-Kolchin2008}
Boylan-Kolchin M.,  Ma C.-P.,   Quataert E.,  2008, \mn@doi [Monthly Notices of
  the Royal Astronomical Society] {10.1111/j.1365-2966.2007.12530.x}, 383, 93

\bibitem[\protect\citeauthoryear{Brough, Couch, Collins, Jarrett, Burke  \&
  Mann}{Brough et~al.}{2008}]{Brough2008}
Brough S.,  Couch W.~J.,  Collins C.~A.,  Jarrett T.,  Burke D.~J.,   Mann
  R.~G.,  2008, \mn@doi [Monthly Notices of the Royal Astronomical Society:
  Letters] {10.1111/j.1745-3933.2008.00442.x}, 385, L103

\bibitem[\protect\citeauthoryear{Bullock, Kolatt, Sigad, Somerville, Kravtsov,
  Klypin, Primack  \& Dekel}{Bullock et~al.}{2001}]{Bullock2001}
Bullock J.~S.,  Kolatt T.~S.,  Sigad Y.,  Somerville R.~S.,  Kravtsov A.~V.,
  Klypin A.~A.,  Primack J.~R.,   Dekel A.,  2001, \mn@doi [Monthly Notices of
  the Royal Astronomical Society] {10.1046/j.1365-8711.2001.04068.x}, 321, 559

\bibitem[\protect\citeauthoryear{Burke, Hilton  \& Collins}{Burke
  et~al.}{2015}]{Burke2015}
Burke C.,  Hilton M.,   Collins C.,  2015, \mn@doi [Monthly Notices of the
  Royal Astronomical Society] {10.1093/mnras/stv450}, 449, 2353

\bibitem[\protect\citeauthoryear{Camps \& Baes}{Camps \&
  Baes}{2015}]{Camps2015}
Camps P.,  Baes M.,  2015, \mn@doi [Astronomy and Computing]
  {10.1016/j.ascom.2014.10.004}, 9, 20

\bibitem[\protect\citeauthoryear{Carilli et~al.,}{Carilli
  et~al.}{2010}]{Carilli2010}
Carilli C.~L.,  et~al., 2010, \mn@doi [The Astrophysical Journal]
  {10.1088/0004-637X/714/2/1407}, 714, 1407

\bibitem[\protect\citeauthoryear{Cerulo, Orellana  \& Covone}{Cerulo
  et~al.}{2019}]{Cerulo2019}
Cerulo P.,  Orellana G.~A.,   Covone G.,  2019, \mn@doi [Monthly Notices of the
  Royal Astronomical Society] {10.1093/mnras/stz1495}

\bibitem[\protect\citeauthoryear{Chandrasekhar}{Chandrasekhar}{1943}]{Chandrasekhar1943a}
Chandrasekhar S.,  1943, \mn@doi [The Astrophysical Journal] {10.1086/144517},
  97, 255

\bibitem[\protect\citeauthoryear{Cimatti, Daddi  \& Renzini}{Cimatti
  et~al.}{2006}]{Cimatti2006}
Cimatti A.,  Daddi E.,   Renzini A.,  2006, \mn@doi [Astronomy {\&}
  Astrophysics] {10.1051/0004-6361:20065155}, 453, L29

\bibitem[\protect\citeauthoryear{Ciotti \& Ostriker}{Ciotti \&
  Ostriker}{1997}]{Ciotti1997}
Ciotti L.,  Ostriker J.~P.,  1997, \mn@doi [The Astrophysical Journal]
  {10.1086/310902}, 487, L105

\bibitem[\protect\citeauthoryear{Collette}{Collette}{2013}]{Collette2013}
Collette A.,  2013, {Python and HDF5}.
O'Reilly

\bibitem[\protect\citeauthoryear{Collins et~al.,}{Collins
  et~al.}{2009}]{Collins2009}
Collins C.~A.,  et~al., 2009, \mn@doi [Nature] {10.1038/nature07865}, 458, 603

\bibitem[\protect\citeauthoryear{Conroy, Wechsler  \& Kravtsov}{Conroy
  et~al.}{2007}]{Conroy2007}
Conroy C.,  Wechsler R.~H.,   Kravtsov A.~V.,  2007, \mn@doi [The Astrophysical
  Journal] {10.1086/521425}, 668, 826

\bibitem[\protect\citeauthoryear{Contini, {De Lucia}, Villalobos  \&
  Borgani}{Contini et~al.}{2014}]{Contini2014}
Contini E.,  {De Lucia} G.,  Villalobos {\'{A}}.,   Borgani S.,  2014, \mn@doi
  [Monthly Notices of the Royal Astronomical Society] {10.1093/mnras/stt2174},
  437, 3787

\bibitem[\protect\citeauthoryear{Cowie \& Binney}{Cowie \&
  Binney}{1977}]{Cowie1977}
Cowie L.~L.,  Binney J.,  1977, \mn@doi [The Astrophysical Journal]
  {10.1086/155406}, 215, 723

\bibitem[\protect\citeauthoryear{Daddi et~al.,}{Daddi et~al.}{2005}]{Daddi2005}
Daddi E.,  et~al., 2005, \mn@doi [The Astrophysical Journal] {10.1086/430104},
  626, 680

\bibitem[\protect\citeauthoryear{Daddi et~al.,}{Daddi et~al.}{2010}]{Daddi2010}
Daddi E.,  et~al., 2010, \mn@doi [The Astrophysical Journal]
  {10.1088/0004-637X/713/1/686}, 713, 686

\bibitem[\protect\citeauthoryear{Dav{\'{e}}, Thompson  \& Hopkins}{Dav{\'{e}}
  et~al.}{2016}]{Dave2016c}
Dav{\'{e}} R.,  Thompson R.,   Hopkins P.~F.,  2016, \mn@doi [Monthly Notices
  of the Royal Astronomical Society] {10.1093/mnras/stw1862}, 462, 3265

\bibitem[\protect\citeauthoryear{Dav{\'{e}}, Rafieferantsoa, Thompson  \&
  Hopkins}{Dav{\'{e}} et~al.}{2017}]{Dave2017}
Dav{\'{e}} R.,  Rafieferantsoa M.~H.,  Thompson R.~J.,   Hopkins P.~F.,  2017,
  \mn@doi [Monthly Notices of the Royal Astronomical Society]
  {10.1093/mnras/stx108}, 467, 115

\bibitem[\protect\citeauthoryear{Dav{\'{e}}, Angl{\'{e}}s-Alc{\'{a}}zar,
  Narayanan, Li, Rafieferantsoa  \& Appleby}{Dav{\'{e}}
  et~al.}{2019}]{Dave2019}
Dav{\'{e}} R.,  Angl{\'{e}}s-Alc{\'{a}}zar D.,  Narayanan D.,  Li Q.,
  Rafieferantsoa M.~H.,   Appleby S.,  2019, \mn@doi [Monthly Notices of the
  Royal Astronomical Society] {10.1093/mnras/stz937}, 486, 2827

\bibitem[\protect\citeauthoryear{{De Lucia} \& Blaizot}{{De Lucia} \&
  Blaizot}{2007}]{DeLucia2007}
{De Lucia} G.,  Blaizot J.,  2007, \mn@doi [Monthly Notices of the Royal
  Astronomical Society] {10.1111/j.1365-2966.2006.11287.x}, 375, 2

\bibitem[\protect\citeauthoryear{{De Lucia}, Springel, White, Croton  \&
  Kauffmann}{{De Lucia} et~al.}{2006}]{DeLucia2006}
{De Lucia} G.,  Springel V.,  White S. D.~M.,  Croton D.,   Kauffmann G.,
  2006, \mn@doi [Monthly Notices of the Royal Astronomical Society]
  {10.1111/j.1365-2966.2005.09879.x}, 366, 499

\bibitem[\protect\citeauthoryear{Diemer \& Kravtsov}{Diemer \&
  Kravtsov}{2015}]{Diemer2015}
Diemer B.,  Kravtsov A.~V.,  2015, \mn@doi [The Astrophysical Journal]
  {10.1088/0004-637X/799/1/108}, 799, 108

\bibitem[\protect\citeauthoryear{Donzelli, Muriel  \& Madrid}{Donzelli
  et~al.}{2011}]{Donzelli2011}
Donzelli C.~J.,  Muriel H.,   Madrid J.~P.,  2011, \mn@doi [The Astrophysical
  Journal Supplement Series] {10.1088/0067-0049/195/2/15}, 195, 15

\bibitem[\protect\citeauthoryear{Douspis, Salvati  \& Aghanim}{Douspis
  et~al.}{2018}]{Douspis2018}
Douspis M.,  Salvati L.,   Aghanim N.,  2018, in 2nd World Summit: Exploring
  the Dark Side of the Universe. 25-29 June. p.~37 (\mn@eprint {arXiv}
  {1901.05289}), \url {http://arxiv.org/abs/1901.05289}

\bibitem[\protect\citeauthoryear{Dubinski}{Dubinski}{1998}]{Dubinski1998}
Dubinski J.,  1998, \mn@doi [The Astrophysical Journal] {10.1086/305901}, 502,
  141

\bibitem[\protect\citeauthoryear{Eggen, Lynden-Bell  \& Sandage}{Eggen
  et~al.}{1962}]{Eggen1962}
Eggen O.~J.,  Lynden-Bell D.,   Sandage A.~R.,  1962, \mn@doi [The
  Astrophysical Journal] {10.1086/147433}, 136, 748

\bibitem[\protect\citeauthoryear{Fabian \& Nulsen}{Fabian \&
  Nulsen}{1977}]{Fabian1977}
Fabian A.~C.,  Nulsen P. E.~J.,  1977, \mn@doi [Monthly Notices of the Royal
  Astronomical Society] {10.1093/mnras/180.3.479}, 180, 479

\bibitem[\protect\citeauthoryear{Faucher-Gigu{\`{e}}re, Lidz, Zaldarriaga  \&
  Hernquist}{Faucher-Gigu{\`{e}}re et~al.}{2009}]{Faucher2009}
Faucher-Gigu{\`{e}}re C.-A.,  Lidz A.,  Zaldarriaga M.,   Hernquist L.,  2009,
  \mn@doi [The Astrophysical Journal] {10.1088/0004-637X/703/2/1416}, 703, 1416

\bibitem[\protect\citeauthoryear{Fontanot, {De Lucia}, Monaco, Somerville  \&
  Santini}{Fontanot et~al.}{2009}]{Fontanot2010}
Fontanot F.,  {De Lucia} G.,  Monaco P.,  Somerville R.~S.,   Santini P.,
  2009, \mn@doi [Monthly Notices of the Royal Astronomical Society]
  {10.1111/j.1365-2966.2009.15058.x}, 397, 1776

\bibitem[\protect\citeauthoryear{Gaburov \& Nitadori}{Gaburov \&
  Nitadori}{2011}]{Gaburov2011}
Gaburov E.,  Nitadori K.,  2011, \mn@doi [Monthly Notices of the Royal
  Astronomical Society] {10.1111/j.1365-2966.2011.18313.x}, 414, 129

\bibitem[\protect\citeauthoryear{Granato, Ragone-Figueroa,
  Dom{\'{i}}nguez-Tenreiro, Obreja, Borgani, {De Lucia}  \& Murante}{Granato
  et~al.}{2015}]{Granato2015}
Granato G.~L.,  Ragone-Figueroa C.,  Dom{\'{i}}nguez-Tenreiro R.,  Obreja A.,
  Borgani S.,  {De Lucia} G.,   Murante G.,  2015, \mn@doi [Monthly Notices of
  the Royal Astronomical Society] {10.1093/mnras/stv676}, 450, 1320

\bibitem[\protect\citeauthoryear{Groves, Dopita, Sutherland, Kewley, Fischera,
  Leitherer, Brandl  \& van Breugel}{Groves et~al.}{2008}]{Groves2008}
Groves B.,  Dopita M.~A.,  Sutherland R.~S.,  Kewley L.~J.,  Fischera J.,
  Leitherer C.,  Brandl B.,   van Breugel W.,  2008, \mn@doi [The Astrophysical
  Journal Supplement Series] {10.1086/528711}, 176, 438

\bibitem[\protect\citeauthoryear{Hernquist}{Hernquist}{1990}]{Hernquist1990}
Hernquist L.,  1990, \mn@doi [The Astrophysical Journal] {10.1086/168845}, 356,
  359

\bibitem[\protect\citeauthoryear{Hernquist}{Hernquist}{1993}]{Hernquist1993}
Hernquist L.,  1993, \mn@doi [The Astrophysical Journal Supplement Series]
  {10.1086/191784}, 86, 389

\bibitem[\protect\citeauthoryear{Higuchi et~al.,}{Higuchi
  et~al.}{2019}]{Higuchi2019}
Higuchi R.,  et~al., 2019, \mn@doi [The Astrophysical Journal]
  {10.3847/1538-4357/ab2192}, 879, 28

\bibitem[\protect\citeauthoryear{Hopkins}{Hopkins}{2015}]{Hopkins2015a}
Hopkins P.~F.,  2015, \mn@doi [Monthly Notices of the Royal Astronomical
  Society] {10.1093/mnras/stv195}, 450, 53

\bibitem[\protect\citeauthoryear{Hopkins, Cox, Younger  \& Hernquist}{Hopkins
  et~al.}{2009}]{Hopkins2009}
Hopkins P.~F.,  Cox T.~J.,  Younger J.~D.,   Hernquist L.,  2009, \mn@doi [The
  Astrophysical Journal] {10.1088/0004-637X/691/2/1168}, 691, 1168

\bibitem[\protect\citeauthoryear{Hopkins et~al.,}{Hopkins
  et~al.}{2018}]{Hopkins2018}
Hopkins P.~F.,  et~al., 2018, \mn@doi [Monthly Notices of the Royal
  Astronomical Society] {10.1093/mnras/sty1690}, 480, 800

\bibitem[\protect\citeauthoryear{Hunter}{Hunter}{2007}]{Hunter2007}
Hunter J.~D.,  2007, \mn@doi [Computing in Science {\&} Engineering]
  {10.1109/MCSE.2007.55}, 9, 90

\bibitem[\protect\citeauthoryear{Ishigaki, Ouchi  \& Harikane}{Ishigaki
  et~al.}{2016}]{Ishigaki2015}
Ishigaki M.,  Ouchi M.,   Harikane Y.,  2016, \mn@doi [The Astrophysical
  Journal] {10.3847/0004-637X/822/1/5}, 822, 5

\bibitem[\protect\citeauthoryear{Ito et~al.,}{Ito et~al.}{2019}]{Ito2019}
Ito K.,  et~al., 2019, \mn@doi [The Astrophysical Journal]
  {10.3847/1538-4357/ab1f0c}, 878, 68

\bibitem[\protect\citeauthoryear{Iwamoto, Brachwitz, Nomoto, Kishimoto, Umeda,
  Hix  \& Thielemann}{Iwamoto et~al.}{1999}]{Iwamoto1999}
Iwamoto K.,  Brachwitz F.,  Nomoto K.,  Kishimoto N.,  Umeda H.,  Hix W.~R.,
  Thielemann F.,  1999, \mn@doi [The Astrophysical Journal Supplement Series]
  {10.1086/313278}, 125, 439

\bibitem[\protect\citeauthoryear{Jiang et~al.,}{Jiang et~al.}{2018}]{Jiang2018}
Jiang L.,  et~al., 2018, \mn@doi [Nature Astronomy]
  {10.1038/s41550-018-0587-9}, 2, 962

\bibitem[\protect\citeauthoryear{Kennicutt}{Kennicutt}{1998}]{Kennicutt1998}
Kennicutt R.~C.,  1998, \mn@doi [The Astrophysical Journal] {10.1086/305588},
  498, 541

\bibitem[\protect\citeauthoryear{Klypin, Trujillo-Gomez  \& Primack}{Klypin
  et~al.}{2011}]{Klypin2011}
Klypin A.~A.,  Trujillo-Gomez S.,   Primack J.,  2011, \mn@doi [The
  Astrophysical Journal] {10.1088/0004-637X/740/2/102}, 740, 102

\bibitem[\protect\citeauthoryear{Klypin, Yepes, Gottl{\"{o}}ber, Prada  \&
  He{\ss}}{Klypin et~al.}{2016}]{Klypin2016}
Klypin A.,  Yepes G.,  Gottl{\"{o}}ber S.,  Prada F.,   He{\ss} S.,  2016,
  \mn@doi [Monthly Notices of the Royal Astronomical Society]
  {10.1093/mnras/stw248}, 457, 4340

\bibitem[\protect\citeauthoryear{Kravtsov, Vikhlinin  \&
  Meshcheryakov}{Kravtsov et~al.}{2018}]{Kravtsov2018}
Kravtsov A.~V.,  Vikhlinin A.~A.,   Meshcheryakov A.~V.,  2018, \mn@doi
  [Astronomy Letters] {10.1134/S1063773717120015}, 44, 8

\bibitem[\protect\citeauthoryear{Krumholz, McKee  \& Tumlinson}{Krumholz
  et~al.}{2009}]{Krumholz2009}
Krumholz M.~R.,  McKee C.~F.,   Tumlinson J.,  2009, \mn@doi [The Astrophysical
  Journal] {10.1088/0004-637X/699/1/850}, 699, 850

\bibitem[\protect\citeauthoryear{Lanson \& Vila}{Lanson \&
  Vila}{2008a}]{Lanson2008a}
Lanson N.,  Vila J.-P.,  2008a, \mn@doi [SIAM Journal on Numerical Analysis]
  {10.1137/S0036142903427718}, 46, 1912

\bibitem[\protect\citeauthoryear{Lanson \& Vila}{Lanson \&
  Vila}{2008b}]{Lanson2008b}
Lanson N.,  Vila J.-P.,  2008b, \mn@doi [SIAM Journal on Numerical Analysis]
  {10.1137/S003614290444739X}, 46, 1935

\bibitem[\protect\citeauthoryear{Laporte \& White}{Laporte \&
  White}{2015}]{Laporte2015}
Laporte C. F.~P.,  White S. D.~M.,  2015, \mn@doi [Monthly Notices of the Royal
  Astronomical Society] {10.1093/mnras/stv112}, 451, 1177

\bibitem[\protect\citeauthoryear{Laporte, White, Naab  \& Gao}{Laporte
  et~al.}{2013}]{Laporte2013}
Laporte C. F.~P.,  White S. D.~M.,  Naab T.,   Gao L.,  2013, \mn@doi [Monthly
  Notices of the Royal Astronomical Society] {10.1093/mnras/stt912}, 435, 901

\bibitem[\protect\citeauthoryear{Lauer, Postman, Strauss, Graves  \&
  Chisari}{Lauer et~al.}{2014}]{Lauer2014}
Lauer T.~R.,  Postman M.,  Strauss M.~A.,  Graves G.~J.,   Chisari N.~E.,
  2014, \mn@doi [The Astrophysical Journal] {10.1088/0004-637X/797/2/82}, 797,
  82

\bibitem[\protect\citeauthoryear{Lavoie et~al.,}{Lavoie
  et~al.}{2016}]{Lavoie2016}
Lavoie S.,  et~al., 2016, \mn@doi [Monthly Notices of the Royal Astronomical
  Society] {10.1093/mnras/stw1906}, 462, 4141

\bibitem[\protect\citeauthoryear{Leitherer et~al.,}{Leitherer
  et~al.}{1999}]{Leitherer1999}
Leitherer C.,  et~al., 1999, \mn@doi [The Astrophysical Journal Supplement
  Series] {10.1086/313233}, 123, 3

\bibitem[\protect\citeauthoryear{Liang, Durier, Babul, Dav{\'{e}}, Oppenheimer,
  Katz, Fardal  \& Quinn}{Liang et~al.}{2016}]{Liang2016}
Liang L.,  Durier F.,  Babul A.,  Dav{\'{e}} R.,  Oppenheimer B.~D.,  Katz N.,
  Fardal M.,   Quinn T.,  2016, \mn@doi [Monthly Notices of the Royal
  Astronomical Society] {10.1093/mnras/stv2840}, 456, 4266

\bibitem[\protect\citeauthoryear{Lidman et~al.,}{Lidman
  et~al.}{2012}]{Lidman2012}
Lidman C.,  et~al., 2012, \mn@doi [Monthly Notices of the Royal Astronomical
  Society] {10.1111/j.1365-2966.2012.21984.x}, 427, 550

\bibitem[\protect\citeauthoryear{Lidman et~al.,}{Lidman
  et~al.}{2013}]{Lidman2013}
Lidman C.,  et~al., 2013, \mn@doi [Monthly Notices of the Royal Astronomical
  Society] {10.1093/mnras/stt777}, 433, 825

\bibitem[\protect\citeauthoryear{Lin \& Mohr}{Lin \& Mohr}{2004}]{Lin2004}
Lin Y.,  Mohr J.~J.,  2004, \mn@doi [The Astrophysical Journal]
  {10.1086/425412}, 617, 879

\bibitem[\protect\citeauthoryear{Loubser, Hoekstra, Babul  \&
  O'Sullivan}{Loubser et~al.}{2018}]{Loubser2018}
Loubser S.~I.,  Hoekstra H.,  Babul A.,   O'Sullivan E.,  2018, \mn@doi
  [Monthly Notices of the Royal Astronomical Society] {10.1093/mnras/sty498},
  477, 335

\bibitem[\protect\citeauthoryear{Martizzi, Hahn, Wu, Evrard, Teyssier  \&
  Wechsler}{Martizzi et~al.}{2016}]{Martizzi2016}
Martizzi D.,  Hahn O.,  Wu H.-Y.,  Evrard A.~E.,  Teyssier R.,   Wechsler
  R.~H.,  2016, \mn@doi [Monthly Notices of the Royal Astronomical Society]
  {10.1093/mnras/stw897}, 459, 4408

\bibitem[\protect\citeauthoryear{McDonald et~al.,}{McDonald
  et~al.}{2016}]{McDonald2016}
McDonald M.,  et~al., 2016, \mn@doi [The Astrophysical Journal]
  {10.3847/0004-637X/817/2/86}, 817, 86

\bibitem[\protect\citeauthoryear{Miller et~al.,}{Miller
  et~al.}{2018}]{Miller2018}
Miller T.~B.,  et~al., 2018, \mn@doi [Nature] {10.1038/s41586-018-0025-2}, 556,
  469

\bibitem[\protect\citeauthoryear{Mo, Mao  \& White}{Mo et~al.}{1998}]{Mo1998}
Mo H.~J.,  Mao S.,   White S. D.~M.,  1998, \mn@doi [Monthly Notices of the
  Royal Astronomical Society] {10.1046/j.1365-8711.1998.01227.x}, 295, 319

\bibitem[\protect\citeauthoryear{Mo, van~den Bosch  \& White}{Mo
  et~al.}{2010}]{Mo2010}
Mo H.,  van~den Bosch F.,   White S.,  2010, {Galaxy Formation and Evolution}.
Cambridge University Press, Cambridge, \mn@doi{10.1017/CBO9780511807244}, \url
  {http://ebooks.cambridge.org/ref/id/CBO9780511807244}

\bibitem[\protect\citeauthoryear{Muzzin et~al.,}{Muzzin
  et~al.}{2009}]{Muzzin2009}
Muzzin A.,  et~al., 2009, \mn@doi [The Astrophysical Journal]
  {10.1088/0004-637X/698/2/1934}, 698, 1934

\bibitem[\protect\citeauthoryear{Narayanan, Bothwell  \& Dav{\'{e}}}{Narayanan
  et~al.}{2012}]{Narayanan2012}
Narayanan D.,  Bothwell M.,   Dav{\'{e}} R.,  2012, \mn@doi [Monthly Notices of
  the Royal Astronomical Society] {10.1111/j.1365-2966.2012.21893.x}, 426, 1178

\bibitem[\protect\citeauthoryear{Nardini, Risaliti, Salvati, Sani, Imanishi,
  Marconi  \& Maiolino}{Nardini et~al.}{2008}]{Nardini2008}
Nardini E.,  Risaliti G.,  Salvati M.,  Sani E.,  Imanishi M.,  Marconi A.,
  Maiolino R.,  2008, \mn@doi [Monthly Notices of the Royal Astronomical
  Society: Letters] {10.1111/j.1745-3933.2008.00450.x}, 385, L130

\bibitem[\protect\citeauthoryear{Navarro, Frenk  \& White}{Navarro
  et~al.}{1997}]{Navarro1997}
Navarro J.~F.,  Frenk C.~S.,   White S. D.~M.,  1997, \mn@doi [The
  Astrophysical Journal] {10.1086/304888}, 490, 493

\bibitem[\protect\citeauthoryear{Neistein, van~den Bosch  \& Dekel}{Neistein
  et~al.}{2006}]{Neistein2006}
Neistein E.,  van~den Bosch F.~C.,   Dekel A.,  2006, \mn@doi [Monthly Notices
  of the Royal Astronomical Society] {10.1111/j.1365-2966.2006.10918.x}, 372,
  933

\bibitem[\protect\citeauthoryear{Nomoto, Tominaga, Umeda, Kobayashi  \&
  Maeda}{Nomoto et~al.}{2006}]{Nomoto2006}
Nomoto K.,  Tominaga N.,  Umeda H.,  Kobayashi C.,   Maeda K.,  2006, \mn@doi
  [Nuclear Physics A] {10.1016/j.nuclphysa.2006.05.008}, 777, 424

\bibitem[\protect\citeauthoryear{Oliphant}{Oliphant}{2006}]{Oliphant2006}
Oliphant T.~E.,  2006, {A guide to NumPy}.
Trelgol Publishing USA

\bibitem[\protect\citeauthoryear{Olsen, Greve, Narayanan, Thompson, Dav{\'{e}},
  Rios  \& Stawinski}{Olsen et~al.}{2017}]{Olsen2017}
Olsen K.,  Greve T.~R.,  Narayanan D.,  Thompson R.,  Dav{\'{e}} R.,  Rios
  L.~N.,   Stawinski S.,  2017, \mn@doi [The Astrophysical Journal]
  {10.3847/1538-4357/aa86b4}, 846, 105

\bibitem[\protect\citeauthoryear{Oppenheimer \& Dav{\'{e}}}{Oppenheimer \&
  Dav{\'{e}}}{2008}]{Oppenheimer2008}
Oppenheimer B.~D.,  Dav{\'{e}} R.,  2008, \mn@doi [Monthly Notices of the Royal
  Astronomical Society] {10.1111/j.1365-2966.2008.13280.x}, 387, 577

\bibitem[\protect\citeauthoryear{Oser, Ostriker, Naab, Johansson  \&
  Burkert}{Oser et~al.}{2010}]{Oser2010}
Oser L.,  Ostriker J.~P.,  Naab T.,  Johansson P.~H.,   Burkert A.,  2010,
  \mn@doi [The Astrophysical Journal] {10.1088/0004-637X/725/2/2312}, 725, 2312

\bibitem[\protect\citeauthoryear{Ostriker \& Tremaine}{Ostriker \&
  Tremaine}{1975}]{Ostriker1975}
Ostriker J.~P.,  Tremaine S.~D.,  1975, \mn@doi [The Astrophysical Journal]
  {10.1086/181992}, 202, L113

\bibitem[\protect\citeauthoryear{Overzier}{Overzier}{2016}]{Overzier2016}
Overzier R.~A.,  2016, \mn@doi [The Astronomy and Astrophysics Review]
  {10.1007/s00159-016-0100-3}, 24, 14

\bibitem[\protect\citeauthoryear{Peebles}{Peebles}{1968}]{Peebles1968}
Peebles P. J.~E.,  1968, \mn@doi [The Astrophysical Journal] {10.1086/149629},
  153, 13

\bibitem[\protect\citeauthoryear{Perez \& Granger}{Perez \&
  Granger}{2007}]{Perez2007}
Perez F.,  Granger B.~E.,  2007, \mn@doi [Computing in Science {\&}
  Engineering] {10.1109/MCSE.2007.53}, 9, 21

\bibitem[\protect\citeauthoryear{Perrin, Soummer, Elliott, Lallo  \&
  Sivaramakrishnan}{Perrin et~al.}{2012}]{Perrin2012}
Perrin M.~D.,  Soummer R.,  Elliott E.~M.,  Lallo M.~D.,   Sivaramakrishnan A.,
   2012, in Clampin M.~C.,  Fazio G.~G.,  MacEwen H.~A.,   Oschmann J.~M.,
  eds,  Vol. 8442, Space Telescopes and Instrumentation 2012: Optical,
  Infrared, and Millimeter Wave. p. 84423D, \mn@doi{10.1117/12.925230}, \url
  {http://proceedings.spiedigitallibrary.org/proceeding.aspx?doi=10.1117/12.925230}

\bibitem[\protect\citeauthoryear{Perrin, Sivaramakrishnan, Lajoie, Elliott,
  Pueyo, Ravindranath  \& Albert}{Perrin et~al.}{2014}]{Perrin2014}
Perrin M.~D.,  Sivaramakrishnan A.,  Lajoie C.-P.,  Elliott E.,  Pueyo L.,
  Ravindranath S.,   Albert L.,  2014, in Oschmann J.~M.,  Clampin M.,  Fazio
  G.~G.,   MacEwen H.~A.,  eds,  Vol. 9143, Space Telescopes and
  Instrumentation 2014: Optical, Infrared, and Millimeter Wave. p. 91433X,
  \mn@doi{10.1117/12.2056689}, \url
  {http://proceedings.spiedigitallibrary.org/proceeding.aspx?doi=10.1117/12.2056689}

\bibitem[\protect\citeauthoryear{Pillepich et~al.,}{Pillepich
  et~al.}{2018}]{Pillepich2018}
Pillepich A.,  et~al., 2018, \mn@doi [Monthly Notices of the Royal Astronomical
  Society] {10.1093/mnras/stx3112}, 475, 648

\bibitem[\protect\citeauthoryear{Pipino, Szabo, Pierpaoli, MacKenzie  \&
  Dong}{Pipino et~al.}{2011}]{Pipino2011}
Pipino A.,  Szabo T.,  Pierpaoli E.,  MacKenzie S.~M.,   Dong F.,  2011,
  \mn@doi [Monthly Notices of the Royal Astronomical Society]
  {10.1111/j.1365-2966.2011.19444.x}, 417, 2817

\bibitem[\protect\citeauthoryear{{Planck Collaboration XVI}}{{Planck
  Collaboration XVI}}{2014}]{PlanckXVI}
{Planck Collaboration XVI} 2014, \mn@doi [Astronomy {\&} Astrophysics]
  {10.1051/0004-6361/201321591}, 571, A16

\bibitem[\protect\citeauthoryear{{Planck Collaboration XX}}{{Planck
  Collaboration XX}}{2014}]{Ade2014}
{Planck Collaboration XX} 2014, \mn@doi [Astronomy {\&} Astrophysics]
  {10.1051/0004-6361/201321521}, 571, A20

\bibitem[\protect\citeauthoryear{Poole, Fardal, Babul, McCarthy, Quinn  \&
  Wadsley}{Poole et~al.}{2006}]{Poole2006}
Poole G.~B.,  Fardal M.~A.,  Babul A.,  McCarthy I.~G.,  Quinn T.,   Wadsley
  J.,  2006, \mn@doi [Monthly Notices of the Royal Astronomical Society]
  {10.1111/j.1365-2966.2006.10916.x}, 373, 881

\bibitem[\protect\citeauthoryear{Postman et~al.,}{Postman
  et~al.}{2012}]{Postman2012}
Postman M.,  et~al., 2012, \mn@doi [The Astrophysical Journal Supplement
  Series] {10.1088/0067-0049/199/2/25}, 199, 25

\bibitem[\protect\citeauthoryear{Ragone-Figueroa, Granato, Ferraro, Murante,
  Biffi, Borgani, Planelles  \& Rasia}{Ragone-Figueroa
  et~al.}{2018}]{Ragone-Figueroa2018}
Ragone-Figueroa C.,  Granato G.~L.,  Ferraro M.~E.,  Murante G.,  Biffi V.,
  Borgani S.,  Planelles S.,   Rasia E.,  2018, \mn@doi [Monthly Notices of the
  Royal Astronomical Society] {10.1093/mnras/sty1639}, 479, 1125

\bibitem[\protect\citeauthoryear{Rennehan, Babul, Hopkins, Dav{\'{e}}  \&
  Moa}{Rennehan et~al.}{2019}]{Rennehan2019}
Rennehan D.,  Babul A.,  Hopkins P.~F.,  Dav{\'{e}} R.,   Moa B.,  2019,
  \mn@doi [Monthly Notices of the Royal Astronomical Society]
  {10.1093/mnras/sty3376}, 483, 3810

\bibitem[\protect\citeauthoryear{Riebe et~al.,}{Riebe et~al.}{2013}]{Riebe2013}
Riebe K.,  et~al., 2013, \mn@doi [Astronomische Nachrichten]
  {10.1002/asna.201211900}, 334, 691

\bibitem[\protect\citeauthoryear{Robertson, Cox, Hernquist, Franx, Hopkins,
  Martini  \& Springel}{Robertson et~al.}{2006}]{Robertson2006}
Robertson B.,  Cox T.~J.,  Hernquist L.,  Franx M.,  Hopkins P.~F.,  Martini
  P.,   Springel V.,  2006, \mn@doi [The Astrophysical Journal]
  {10.1086/500360}, 641, 21

\bibitem[\protect\citeauthoryear{Romer, Viana, Liddle  \& Mann}{Romer
  et~al.}{2001}]{Romer2001}
Romer A.~K.,  Viana P. T.~P.,  Liddle A.~R.,   Mann R.~G.,  2001, \mn@doi [The
  Astrophysical Journal] {10.1086/318382}, 547, 594

\bibitem[\protect\citeauthoryear{Ruszkowski \& Springel}{Ruszkowski \&
  Springel}{2009}]{Ruszkowski2009}
Ruszkowski M.,  Springel V.,  2009, \mn@doi [The Astrophysical Journal]
  {10.1088/0004-637X/696/2/1094}, 696, 1094

\bibitem[\protect\citeauthoryear{Sahl{\'{e}}n et~al.,}{Sahl{\'{e}}n
  et~al.}{2009}]{Sahlen2009}
Sahl{\'{e}}n M.,  et~al., 2009, \mn@doi [Monthly Notices of the Royal
  Astronomical Society] {10.1111/j.1365-2966.2009.14923.x}, 397, 577

\bibitem[\protect\citeauthoryear{Salvati, Douspis  \& Aghanim}{Salvati
  et~al.}{2018}]{Salvati2018}
Salvati L.,  Douspis M.,   Aghanim N.,  2018, \mn@doi [Astronomy {\&}
  Astrophysics] {10.1051/0004-6361/201731990}, 614, A13

\bibitem[\protect\citeauthoryear{Sandage}{Sandage}{1976}]{Sandage1976}
Sandage A.,  1976, \mn@doi [The Astrophysical Journal] {10.1086/154315}, 205, 6

\bibitem[\protect\citeauthoryear{Scannapieco \& Bildsten}{Scannapieco \&
  Bildsten}{2005}]{Scannapieco2005}
Scannapieco E.,  Bildsten L.,  2005, \mn@doi [The Astrophysical Journal]
  {10.1086/452632}, 629, L85

\bibitem[\protect\citeauthoryear{Schaye et~al.,}{Schaye
  et~al.}{2015}]{Schaye2014}
Schaye J.,  et~al., 2015, \mn@doi [Monthly Notices of the Royal Astronomical
  Society] {10.1093/mnras/stu2058}, 446, 521

\bibitem[\protect\citeauthoryear{Shankar et~al.,}{Shankar
  et~al.}{2015}]{Shankar2015}
Shankar F.,  et~al., 2015, \mn@doi [The Astrophysical Journal]
  {10.1088/0004-637X/802/2/73}, 802, 73

\bibitem[\protect\citeauthoryear{Silk \& Rees}{Silk \& Rees}{1998}]{Silk1998}
Silk J.,  Rees M.~J.,  1998, Monthly Notices of the Royal Astronomical Society,
  324, 128

\bibitem[\protect\citeauthoryear{Smith et~al.,}{Smith et~al.}{2017}]{Smith2017}
Smith B.~D.,  et~al., 2017, \mn@doi [Monthly Notices of the Royal Astronomical
  Society] {10.1093/mnras/stw3291}, 466, 2217

\bibitem[\protect\citeauthoryear{Springel}{Springel}{2000}]{Springel2000}
Springel V.,  2000, \mn@doi [Monthly Notices of the Royal Astronomical Society]
  {10.1046/j.1365-8711.2000.03187.x}, 312, 859

\bibitem[\protect\citeauthoryear{Springel \& White}{Springel \&
  White}{1999}]{Springel1999}
Springel V.,  White S. D.~M.,  1999, \mn@doi [Monthly Notices of the Royal
  Astronomical Society] {10.1046/j.1365-8711.1999.02613.x}, 307, 162

\bibitem[\protect\citeauthoryear{Springel, {Di Matteo}  \& Hernquist}{Springel
  et~al.}{2005}]{Springel2005b}
Springel V.,  {Di Matteo} T.,   Hernquist L.,  2005, \mn@doi [Monthly Notices
  of the Royal Astronomical Society] {10.1111/j.1365-2966.2005.09238.x}, 361,
  776

\bibitem[\protect\citeauthoryear{Stott et~al.,}{Stott et~al.}{2010}]{Stott2010}
Stott J.~P.,  et~al., 2010, \mn@doi [Astrophysical Journal]
  {10.1088/0004-637X/718/1/23}, 718, 23

\bibitem[\protect\citeauthoryear{Stott, Collins, Burke, Hamilton-Morris  \&
  Smith}{Stott et~al.}{2011}]{Stott2011}
Stott J.~P.,  Collins C.~A.,  Burke C.,  Hamilton-Morris V.,   Smith G.~P.,
  2011, \mn@doi [Monthly Notices of the Royal Astronomical Society]
  {10.1111/j.1365-2966.2011.18404.x}, 414, 445

\bibitem[\protect\citeauthoryear{Tabor \& Binney}{Tabor \&
  Binney}{1993}]{Tabor1993}
Tabor G.,  Binney J.,  1993, \mn@doi [Monthly Notices of the Royal Astronomical
  Society] {10.1093/mnras/263.2.323}, 263, 323

\bibitem[\protect\citeauthoryear{Tacconi et~al.,}{Tacconi
  et~al.}{2010}]{Tacconi2010}
Tacconi L.~J.,  et~al., 2010, \mn@doi [Nature] {10.1038/nature08773}, 463, 781

\bibitem[\protect\citeauthoryear{Tacconi et~al.,}{Tacconi
  et~al.}{2013}]{Tacconi2013}
Tacconi L.~J.,  et~al., 2013, \mn@doi [The Astrophysical Journal]
  {10.1088/0004-637X/768/1/74}, 768, 74

\bibitem[\protect\citeauthoryear{Tadaki et~al.,}{Tadaki
  et~al.}{2019}]{Tadaki2019}
Tadaki K.-i.,  et~al., 2019, \mn@doi [Publications of the Astronomical Society
  of Japan] {10.1093/pasj/psz005}, 71, 1

\bibitem[\protect\citeauthoryear{Teyssier, Moore, Martizzi, Dubois  \&
  Mayer}{Teyssier et~al.}{2011}]{Teyssier2011}
Teyssier R.,  Moore B.,  Martizzi D.,  Dubois Y.,   Mayer L.,  2011, \mn@doi
  [Monthly Notices of the Royal Astronomical Society]
  {10.1111/j.1365-2966.2011.18399.x}, 414, 195

\bibitem[\protect\citeauthoryear{Tremaine \& Richstone}{Tremaine \&
  Richstone}{1977}]{Tremaine1977}
Tremaine S.~D.,  Richstone D.~O.,  1977, \mn@doi [The Astrophysical Journal]
  {10.1086/155049}, 212, 311

\bibitem[\protect\citeauthoryear{Tremmel, Governato, Volonteri  \&
  Quinn}{Tremmel et~al.}{2015}]{Tremmel2015a}
Tremmel M.,  Governato F.,  Volonteri M.,   Quinn T.~R.,  2015, \mn@doi
  [Monthly Notices of the Royal Astronomical Society] {10.1093/mnras/stv1060},
  451, 1868

\bibitem[\protect\citeauthoryear{Trudeau et~al.,}{Trudeau
  et~al.}{2019}]{Trudeau2019}
Trudeau A.,  et~al., 2019, \mn@doi [Monthly Notices of the Royal Astronomical
  Society] {10.1093/mnras/stz1364}, 487, 1210

\bibitem[\protect\citeauthoryear{Vitvitska, Klypin, Kravtsov, Wechsler, Primack
   \& Bullock}{Vitvitska et~al.}{2002}]{Vitvitska2002}
Vitvitska M.,  Klypin A.~A.,  Kravtsov A.~V.,  Wechsler R.~H.,  Primack J.~R.,
   Bullock J.~S.,  2002, \mn@doi [The Astrophysical Journal] {10.1086/344361},
  581, 799

\bibitem[\protect\citeauthoryear{Webb et~al.,}{Webb et~al.}{2015}]{Webb2015}
Webb T. M.~A.,  et~al., 2015, \mn@doi [The Astrophysical Journal]
  {10.1088/0004-637X/814/2/96}, 814, 96

\bibitem[\protect\citeauthoryear{Webb et~al.,}{Webb et~al.}{2017}]{Webb2017}
Webb T. M.~A.,  et~al., 2017, \mn@doi [The Astrophysical Journal]
  {10.3847/2041-8213/aa7749}, 844, L17

\bibitem[\protect\citeauthoryear{Whiley et~al.,}{Whiley
  et~al.}{2008}]{Whiley2008}
Whiley I.~M.,  et~al., 2008, \mn@doi [Monthly Notices of the Royal Astronomical
  Society] {10.1111/j.1365-2966.2008.13324.x}, 387, 1253

\bibitem[\protect\citeauthoryear{White}{White}{1976}]{White1976}
White S. D.~M.,  1976, \mn@doi [Monthly Notices of the Royal Astronomical
  Society] {10.1093/mnras/174.1.19}, 174, 19

\bibitem[\protect\citeauthoryear{Wilson et~al.,}{Wilson
  et~al.}{2009}]{Wilson2009}
Wilson G.,  et~al., 2009, \mn@doi [The Astrophysical Journal]
  {10.1088/0004-637X/698/2/1943}, 698, 1943

\bibitem[\protect\citeauthoryear{Zubko, Dwek  \& Arendt}{Zubko
  et~al.}{2004}]{Zubko2004}
Zubko V.,  Dwek E.,   Arendt R.~G.,  2004, \mn@doi [The Astrophysical Journal
  Supplement Series] {10.1086/382351}, 152, 211

\bibitem[\protect\citeauthoryear{van~der Walt, Colbert  \& Varoquaux}{van~der
  Walt et~al.}{2011}]{VanderWalt2011}
van~der Walt S.,  Colbert S.~C.,   Varoquaux G.,  2011, \mn@doi [Computing in
  Science {\&} Engineering] {10.1109/MCSE.2011.37}, 13, 22

\bibitem[\protect\citeauthoryear{van~der Wel, Holden, Zirm, Franx, Rettura,
  Illingworth  \& Ford}{van~der Wel et~al.}{2008}]{VanderWel2008}
van~der Wel A.,  Holden B.~P.,  Zirm A.~W.,  Franx M.,  Rettura A.,
  Illingworth G.~D.,   Ford H.~C.,  2008, \mn@doi [The Astrophysical Journal]
  {10.1086/592267}, 688, 48

\makeatother
\end{thebibliography}



\appendix

\section{JWST observing strategy}
\label{app:mock_observations}

To generate mock NIRCam images, we select an optimal observing strategy for our synthetic SPT2349-56 system assuming 10 ks of observing time.  We used the MUTLIACCUM MEDIUM 8 observing read-mode using Module B, in which each 10$\times$1ks integration is divided non-destructively into 10$\times$100s groups with 8$\times$10s frames/group and 2$\times$10s drop frames/group. This MULTIACCUM "up-the-ramp" observing strategy enables cosmic ray rejection, reduces the readout noise (roughly by the square root of the total number of frames) and increases the dynamic range by preventing saturation by bright sources. We generate point spread function (PSF) images for this observing strategy in each band pass using \pkg{pynrc} and the \pkg{WebbPSF} package \citep{Perrin2012, Perrin2014} and convolve with the idealised image. The $1\sigma$ AB surface brightness sensitivities determined by \pkg{pynrc} for the NIRCam detector with this observing strategy were $m_{\mathrm{AB,band}} = {27.38,27.62,28.52,28.65}$ AB $\mathrm{mag} \, \mathrm{arcsec}^{-2}$ for {F150W, F200W, F277W, F356W}.  These sensitivity estimates are based on characterization data for the detectors including readout and 1/f noise, dark current, and background levels equal to 1.2 times the minimum zodiacal light background.  The current development version of \pkg{pynrc} does not allow us to convert between calibrated and non-calibrated images (i.e., AB zero points, effective gain and exposure time), so we do not incorporate Poisson shot noise from the source light.  For now, we naively model the total noise contributions as a single Gaussian process with standard deviation equal to the reported sensitivity in each band pass using our observing strategy. 


\bsp	
\label{lastpage}
\end{document}